\documentclass[twocolumn]{aastex6}

\usepackage{graphicx,epstopdf,amsmath}
\usepackage{import}

\usepackage{color}

\newcommand{\phpl}{PhPl}
\newcommand{\sci}{Sci}

\newcommand{\cophc}{CoPhC}

\newcommand{\soph}{SoPh}

\newcommand{\jgra}{JGRA}

\newcommand{\prlet}{PhRvL}
\newcommand{\raa}{RAA}

\shorttitle{A Breakout Model for Solar Coronal Jets with Filaments}
\shortauthors{Wyper, DeVore, \& Antiochos}

\begin{document}

\title{A Breakout Model for Solar Coronal Jets with Filaments}


\author{P.~F.~Wyper} 
\affil{Department of Mathematical Sciences, Durham University, Durham, DH1 3LE, UK}
\email{peter.f.wyper@durham.ac.uk}

\author{C.~R.~DeVore} 
\affil{Heliophysics Science Division, NASA Goddard Space Flight Center, 8800 Greenbelt Rd, Greenbelt, MD 20771}
\email{c.richard.devore@nasa.gov}

\author{S.~K.~Antiochos} 
\affil{Heliophysics Science Division, NASA Goddard Space Flight Center, 8800 Greenbelt Rd, Greenbelt, MD 20771}
\email{spiro.antiochos@nasa.gov}

\begin{abstract}
Recent observations have revealed that many solar coronal jets involve the eruption of miniature versions of large-scale filaments. Such ``mini-filaments" are observed to form along the polarity inversion lines of strong, magnetically bipolar regions embedded in open (or distantly closing) unipolar field. During the generation of the jet, the filament becomes unstable and erupts. Recently we described a model for these mini-filament jets, in which the well-known magnetic-breakout mechanism for large-scale coronal mass ejections is extended to these smaller events. In this work we use three-dimensional magnetohydrodynamic simulations to study in detail three realisations of the model. We show that the breakout-jet generation mechanism is robust and that different realisations of the model can explain different observational features. The results are discussed in relation to recent observations and previous jet models.
\end{abstract}



\section{Introduction}
Coronal jets are transient, collimated ejections of plasma launched from low in the solar atmosphere outwards along the ambient magnetic field of the corona. Jets occur prolifically across the solar surface, most notably within coronal holes and around the periphery of active regions \citep{Shimojo1996,Savcheva2007}. They are observed in X-rays \citep[e.g.][]{Shimojo1996,Cirtain2007} and at a variety of extreme ultraviolet (EUV) wavelengths \citep[e.g.][]{Nistico2009,Zhang2016b}, reflecting the fact that some jets possess both hot and cool (relative to the ambient corona) components. Some jets are energetic enough to reach the heliosphere and become visible as jet-like CMEs in white-light coronagraphs \citep[e.g.][]{Wang1998,Patsourakos2008,Hong2011,Moore2015}. From X-ray observations, \citet{Savcheva2007} found that a sample of around $100$ jets had typical lifetimes of around $10$ min, lengths on the order of $50$ Mm, widths of around $8$ Mm, and bulk outflow velocities of around $200 \,\text{km}\,\text{s}^{-1}$. {For a comprehensive review of jet observations, morphologies, and previous numerical modeling see \citet{Raouafi2016}.}

\begin{figure}
\centering
\includegraphics[width=0.45\textwidth]{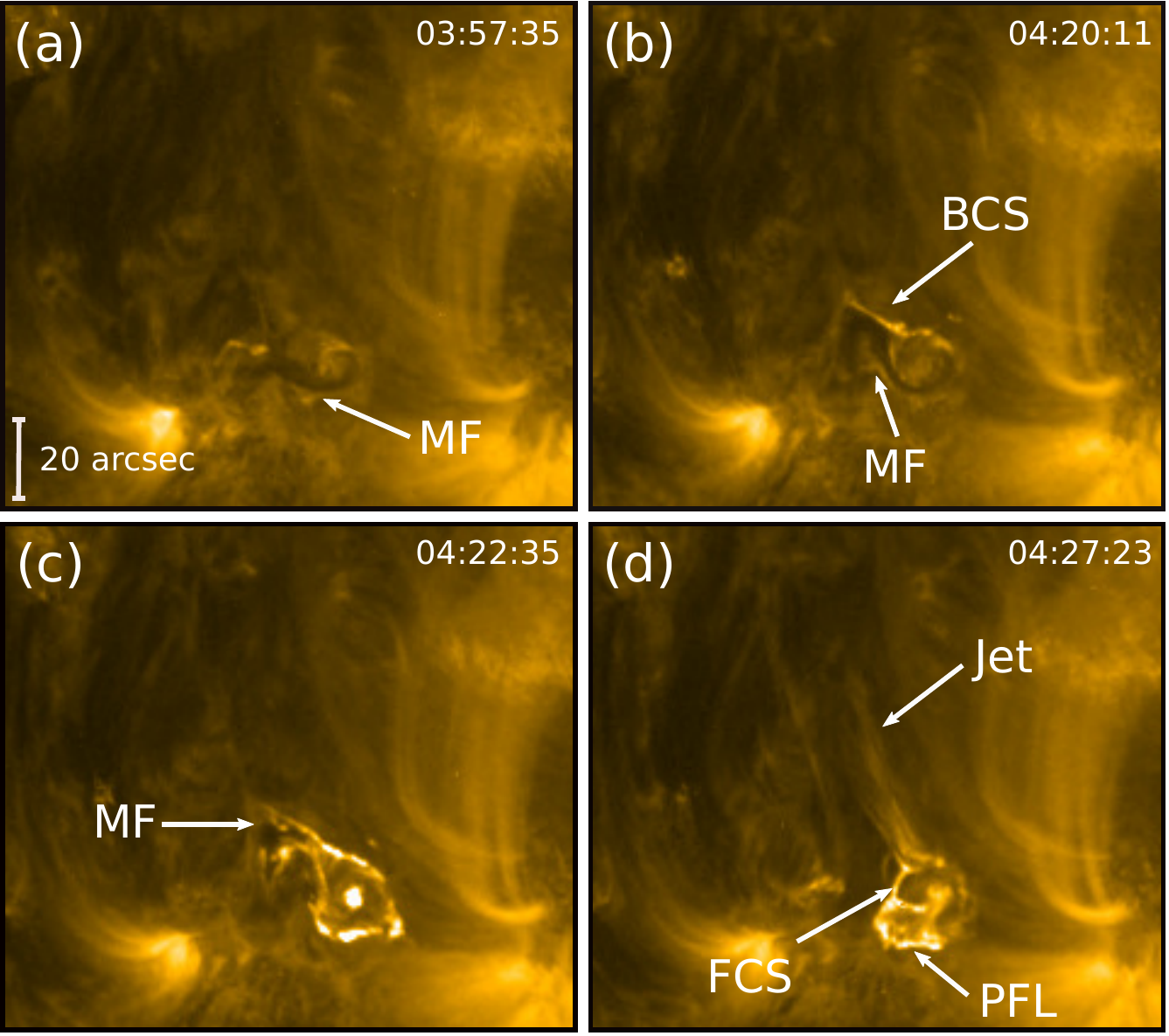}
\caption{An example of a mini-filament jet. MF = mini-filament, BCS = breakout current sheet, FCS = flare current sheet and PFL = post-flare loops. See text for details.}
\label{fig:obs}
\end{figure}

Common to all jets is an impulsive release of energy as the jet is launched, accompanied by the formation of hot magnetic loops off to one side of the jet base. \citet{Shibata1992} proposed that the plasma jet and the bright loops (also called the jet bright point) could be explained by the emergence of a small bipole into a unipolar region of open (or distantly closing) magnetic field. External reconnection between the emerging flux and the ambient field would produce a jet of plasma and form a new set of hot, reconnected loops. Numerous numerical experiments have tested this idea and shown that such a jet outflow with loops can be realised from that scenario \citep[e.g.][]{Yokoyama1995,Yokoyama1996,Miyagoshi2003,Miyagoshi2004,Archontis2005,Galsgaard2005,Moreno-Insertis2008,Gontikakis2009,Archontis2010}.

Many jets exhibit the classic inverted-Y, or Eiffel-tower, shape consistent with the Shibata picture. However, a large proportion of jets instead have a broad jet spire that often exhibits strong helical motion \citep[e.g.][]{Patsourakos2008,Nistico2009,Shen2011,Hong2013,Moore2015}. {The prevalence of different jet morphologies was studied by \citet{Nistico2009} using EUV observations from the {\it Solar TErrestrial RElations Observatory}. From a sample of 79 jets, their study classified 31 as exhibiting observable helical motions; structurally, 37 were of Eiffel tower-type, 12 as the similar lambda-type, 5 resembled miniature coronal mass ejections (CMEs), and the remaining 25 were ambiguous.}

{Some authors have noted that most jets with helical motions seem to involve the eruption, or blowing out, of the bipole region in a form reminiscent of mini-CMEs} \citep{Nistico2009,Innes2009,Innes2010,Moore2010,Raouafi2010}. \citet{Moore2010} suggested that such ``blowout'' jets could be explained by an extension of the Shibata ``standard'' jet in which the emerging bipole becomes unstable, a section of it erupts, and flare-like loops form underneath to create the jet bright point. Several numerical experiments have now replicated this behaviour when the flux emergence continues over a sufficiently long period \citep[e.g.][]{Archontis2013,Moreno-Insertis2013,Fang2014}. 

However, recent observations suggest that flux emergence is not the fundamental driver of all coronal jets. In jets where magnetogram data are available, {it is often observed that little or no flux emergence occurs leading up to or during the jet} \citep[e.g.][]{Chandrashekhar2014,Hong2016}. More typically, flux is actually cancelling at the base of the jet \citep[e.g.][]{Chae1999,Liu2011,Hong2011,Young2014a,Young2014b,Adams2014,Panesar2016}. Therefore, while flux emergence is highly likely to account for {the generation of} some coronal jets, it seems improbable that it explains all such events. 

A particular challenge to the the flux-emergence model is posed by the relatively recent identification in many jets of small filament-like structures that are invisible in X-rays, but can be seen in wavelengths associated with cooler plasma, such as EUV and $\text{H}\alpha$ \citep[e.g.][]{Zheng2012,Sterling2015,Hong2016,Zhang2016b}. These ``mini-filaments'' are observed along and above the polarity inversion lines (PILs) of pre-existing strong bipoles \citep[e.g.][]{Chae1999,Hong2011,Hong2014,Hong2016,Zheng2012,Adams2014,Panesar2016,Zhang2016b}. They contain cool, dense plasma (relative to the ambient corona) and resemble the large-scale filaments that erupt as CMEs, leaving behind arcades of bright flare loops. In an apparently similar manner, these mini-filaments erupt and leave behind the loops of the jet bright point, as the jet itself propagates away through the corona. The ejection of cool filament material alongside hot plasma heated by reconnection may explain the often observed simultaneous occurrence of hot jets and their cooler counterparts, surges \citep[e.g.][]{Canfield1996}. 

{An example of a mini-filament jet in a quiet-Sun region \cite[one of several studied by][]{Panesar2016} is shown in Figure \ref{fig:obs} depicting the typical phases of evolution. The EUV images were taken with the Atmospheric Imaging Assembly aboard the {\it Solar Dynamics Observatory} at 171 \AA \, ($T \approx 0.7\, \text{MK}$) on 2012 November 13 and were rendered using Helioviewer (www.helioviewer.org). Prior to the jet a pre-existing, dark mini-filament is present (a). The overlying structure then slowly begins to rise as a bright linear feature, which we interpret as showing the breakout current layer (see later sections) forms above it (b). An explosive change in evolution occurs as the dark mini-filament reaches the linear feature (c), after which an untwisting jet is launched (d). Concurrent with the launching of the jet is the formation of post-flare loops where the mini-filament was present initially. In addition, a second bright linear feature forms, connecting the top of the loops and the untwisting jet curtain, which we interpret as showing the flare current layer (see also later sections).}

Recently, \citet{Sterling2015} examined $20$ randomly selected jets and found that all involved mini-filaments. The largest were comparable in size to the closed-field region and erupted to form broad blowout jets with a strongly rotating spire. The smallest were associated with jets exhibiting the classic inverted-Y shape. Sterling et al.\ concluded that all jets stem from the eruption of mini-filaments and that flux emergence plays little or no role in jet generation. 

To date, only one family of jet models has generated jets without flux emergence. In the model pioneered by Pariat and collaborators \citep{Pariat2009,Pariat2010,Pariat2015,Pariat2016,Dalmasse2012,Wyper2016,Wyper2016b,Karpen2017}, a single strong-polarity region embedded in an ambient field of opposite sign is slowly rotated to store magnetic free energy and helicity in the corona. Eventually, a kink-like instability induces explosive interchange reconnection between the rotationally sheared closed field and the external, unsheared open field. The resulting jets have a strong rotational component, consistent with blowout jets, along with a broad spire and enhanced density in the jet over and above the coronal background density. These kink-induced jets are one realisation of the ``sweeping magnetic twist" jet mechanism proposed by \citet{Shibata1986}, whereby the transfer of twist/shear by reconnection from closed to open field lines drives a rotating jet, as the twist propagates along the ambient field as a nonlinear Alfv\'{e}n wave. In their original formulation of the model, Shibata \& Uchida envisaged this twist to be stored within a filament, as recent observations have now revealed. However, the simulations cited above lack an internal magnetic structure that would support a filament and do not readily explain the observed asymmetry of the hot reconnected loops at the base following the jet generation.

Motivated by the observations of \citet{Sterling2015}, we conducted a high resolution MHD simulation \citep{Wyper2017} showing how the ``magnetic breakout" mechanism \citep{Antiochos1999} for large-scale CMEs is universal, also explaining small-scale coronal jets involving mini-filaments. In these jets, free energy is stored in the filament channel along the PIL of a bipole embedded in an open ambient magnetic field; the filament subsequently erupts to form a blowout-like untwisting jet. In this paper, we present the details of this model for coronal jets and demonstrate three realisations with ambient fields of different inclinations. The results in each case show that the filament-channel field erupts following the onset of a magnetic-breakout reconnection process at the overlying coronal null point, exactly analogous to the breakout mechanism for large-scale filament eruptions and CMEs \citep{Antiochos1999,MacNeice2004,Phillips2005,DeVore2008,Lynch2008,Lynch2009,Karpen2012,Masson2013}. The slow rise of the filament/flux rope prior to the breakout phase, the subsequent fast flare-like reconnection, and the generation of the impulsive untwisting jet are all consistent with the observations of jets associated with mini-filament eruptions {(Fig.\ \ref{fig:obs}).}

Section \ref{sec:setup} describes the setup of the simulations and the details of the model. In \S \ref{sec:mechanism}, we present a schematic outline of the jet mechanism and the different phases of the evolution. Section \ref{sec:overview} describes the energies and general morphologies of the simulated jets, whilst in \S \ref{sec:phases} we discuss the various evolutionary phases and the differences between them in detail. In \S \ref{sec:fluxes}, we analyze the Poynting and kinetic-energy fluxes transferred by the jet into the corona. Finally, we discuss our findings in \S \ref{sec:discussion}.

\begin{table}[ht]
\centering 
\caption{Dipole parameters.} 
\begin{tabular}{c c c c c c} 
\hline\hline 
i & $b_{i}$ & $x_{i}$ & $y_{i}$ & $z_{i}$ \\ [0.5ex] 
\hline 
1 & $\phantom{-}6.0$ & $-1.0$ & $-0.5$ & $-1.0$ \\ 
2 & $\phantom{-}6.0$ & $-1.0$ & $-0.5$ & $-0.5$ \\
3 & $\phantom{-}6.0$ & $-1.0$ & $-0.5$ & $\phantom{-}0.0$ \\
4 & $\phantom{-}6.0$ & $-1.0$ & $-0.5$ & $\phantom{-}0.5$ \\
5 & $\phantom{-}6.0$ & $-1.0$ & $-0.5$ & $\phantom{-}1.0$ \\
6 & $\phantom{-}6.0$ & $-1.0$ & $\phantom{-}0.0$ & $\phantom{-}0.0$ \\
7 & $\phantom{-}6.0$ & $-1.0$ & $\phantom{-}0.0$ & $\phantom{-}1.0$ \\
8 & $\phantom{-}6.0$ & $-1.0$ & $\phantom{-}0.0$ & $-1.0$ \\
9 & $-5.3$ & $\phantom{-}1.0$ & $\phantom{-}1.5$ & $-1.0$ \\
10 & $-5.3$ & $-1.0$ & $\phantom{-}1.5$ & $-0.5$ \\
11 & $-5.3$ & $-1.0$ & $\phantom{-}1.5$ & $\phantom{-}0.0$ \\
12 & $-5.3$ & $-1.0$ & $\phantom{-}1.5$ & $\phantom{-}0.5$ \\
13 & $-5.3$ & $-1.0$ & $\phantom{-}1.5$ & $\phantom{-}1.0$ \\
14 & $-5.3$ & $-1.0$ & $\phantom{-}1.0$ & $\phantom{-}0.0$ \\
15 & $-5.3$ & $-1.0$ & $\phantom{-}1.0$ & $\phantom{-}1.0$ \\
16 & $-5.3$ & $-1.0$ & $\phantom{-}1.0$ & $-1.0$ \\
\hline 
\end{tabular}
\label{table:bipoles} 
\end{table}

\begin{figure*}
\centering
\includegraphics[width=0.7\textwidth]{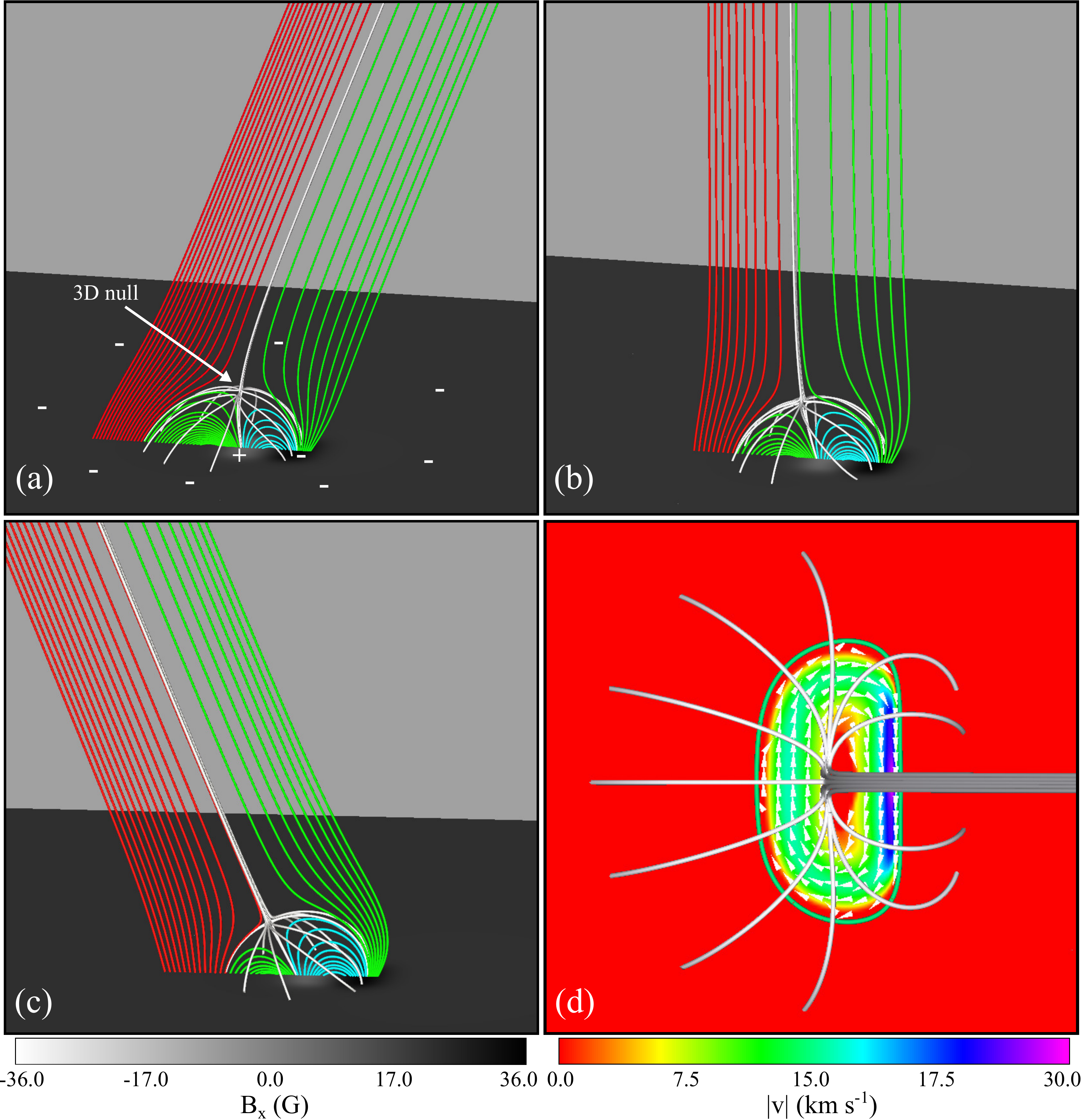}
\caption{The initial potential magnetic field in the three simulations with background-field tilt angles (a) $\theta = +22^\circ$, (b) $\theta = 0^\circ$, and (c) $\theta = -22^\circ$. The field is comprised of the domed fan plane and spine lines (silver field lines) of a 3D coronal null point above the parasitic polarity of a bipolar photospheric flux distribution. (d) Driving flows tangential to the photospheric boundary follow the contours of the positive parasitic polarity and are shown for $\theta = +22^\circ$. Note the increased flow speed near the polarity inversion line (green contour of $B_x = 0$) in the centre of the bipolar distribution.}
\label{fig:setup}
\end{figure*}

\section{Simulation Setup}
\label{sec:setup}
Observations suggest that the mini-filament erupting in conjunction with a coronal jet is confined beneath closed small-scale coronal loops, which in turn are embedded in the open field along which the jet subsequently propagates. This is highly suggestive of a filament channel forming and subsequently erupting from the closed field beneath an overlying coronal null point. To model this process, we adopt an initially potential field with a compact bipolar structure on the solar surface (where the filament channel will be formed) embedded within a uniform inclined background field. Figure \ref{fig:setup} shows the field in each of the simulations that we performed. The strong bipolar field near the surface creates a confining strapping field (cyan field lines) above the polarity inversion line (PIL) beneath the separatrix of the 3D null (silver field lines). The field is constructed by superposing $16$ vertically oriented sub-photospheric dipoles and the uniform background field,
\begin{align}
\mathbf{B} &= \left(c_{1}\cos\theta,c_{1}\sin\theta,0\right) + \sum_{i=1,16} \boldsymbol{\nabla} \times \mathbf{A}_{i},\\
\mathbf{A}_{i} &= \frac{b_{i}x_{i}^{3}}{2\left[ x_{i}'^{2} + (y_{i}'-y_{c})^{2} + z_{i}'^{2} \right]^{3/2}}\nonumber\\
               &\times \left[ -z_{i}' \hat{\mathbf{y}} + (y_{i}'-y_{c}) \hat{\mathbf{z}} \right],
\end{align}
where $c_{1} = -1.077$, $\theta$ is the angle of the field clockwise from the vertical, and $x_{i}' = x-x_{i}$, $y_{i}' = y-y_{i}$, and $z_{i}' = z-z_{i}$. The values for $b_{i}, x_{i}, y_{i},$ and $z_{i}$ are given in Table \ref{table:bipoles}. The photosphere is located at $x=0$. In the three simulations, we set $\theta = -22^\circ, 0^\circ,$ and $+22^\circ$ (corresponding to Figs.\ \ref{fig:setup}(a), (b) and (c) respectively). The coordinate $y_{c}$ determines the position along the $y$-axis where the bipolar region is situated; we used $y_{c} = -5.0, -0.5,$ and $-5.0$ for $\theta = -22^\circ, 0^\circ,$ and $+22^\circ$, respectively. The peak field strength in the parasitic polarity in each configuration is $B \approx 17$. The width of the separatrix dome at the photosphere varies from $w \approx 5$ at its narrowest to $w \approx 7$ at its widest, giving an average width of $w \approx 6$. The height of the null varies from $h \approx 1.7$ for $\theta = -22^\circ$ to $ h\approx 2.0$ for $\theta = +22^\circ$. 

{For maximum generality, we solved the equations in non-dimensional form. For purposes of direct comparison to observations, we can introduce scaling factors typical of the corona. {For the set of equations solved in the simulation (given below)}, fixing a typical length scale ($L_{s}$), plasma density ($\rho_{s}$), and magnetic field strength ($B_{s}$) is sufficient to fully define the scale values of other variables. We have
\begin{align}
V_s &= \frac{B_s}{\sqrt{\rho_s}}, \quad t_s = \frac{L_s}{V_s},\\
E_s &= B_s^2 L_s^3, \quad P_s = B_s^2,
\end{align}
where $V_{s}$ scales the velocity, $t_{s}$ the time, $E_{s}$ the total energy, and $P_{s}$ the pressure (and, subsequently, the temperature through the ideal gas law). Choosing $L_{s} = 4\times10^{8} \,\text{cm}$, $B_{s} = 2\,\text{G}$, and $\rho_{s} = 4\times 10^{-16} \,\text{g}\,\text{cm}^{-3}$ give $V_{s} = 1000$ km s$^{-1}$, $t_{s} = 4$ s, $E_{s} = 2.56 \times 10^{26}$ erg, and $P_{s} = 4$ dyn cm$^{-2}$. The scaled average width of the separatrix dome in each case becomes ${\bar w} \approx 24$ Mm, and the scaled peak field strength within the parasitic polarity becomes ${\bar B} \approx 34$ G. In the results presented below, we use these scalings to convert our non-dimensional numerical results to solar values. Note, however, that the quoted values can be modified by redefining any of our baseline scale parameters $L_{s}$, $\rho_{s}$, and $B_{s}$ to apply the results to a particular observed jet.}


The filament channel is created by prescribing a photospheric flow that follows the contours of $B_{x}$ within the positive (parasitic) polarity patch of the large-scale bipole. The spatial dependence of the velocity pattern is calculated from \citep{Wyper2016}
\begin{align}
\mathbf{v}_{\perp} &= v_{0}g(B_x)\hat{\mathbf{x}}\times \boldsymbol{\nabla} B_{x},\\
g(B_x) &= k_{B}\frac{B_{r}-B_{l}}{B_{x}}\tanh\left(k_{B}\frac{B_{x}-B_{l}}{B_{r}-B_{l}}\right), B_l \le B_x \le B_r,\\
       &= 0, {\rm otherwise}\nonumber
\label{eq:vdef}
\end{align}
where $B_x$ is the spatially varying vertical field component, and $B_l$, $B_r$, $k_B$, and $v_0$ are fixed constants set to $0.8$, $15.0$, $4.0$, and $1.0\times 10^{-4}$, respectively. Figure \ref{fig:setup} (d) shows the driving flow for $\theta = +22^\circ$ (the photospheric field distribution, and hence the velocity profile, vary little about $y = y_c$ among the three experiments). The high gradient in $B_{x}$ across the PIL at the centre of the bipole generates the fastest flows at this location (purple strip on the right hand side) and helps to form the filament channel there. The driving is reduced to zero a small distance from the centre of the positive polarity to minimize the perturbation applied to the inner spine of the null. The peak driving speed for this velocity profile is $v_{\perp} \approx 0.03$ {($30 \text{ km}\,\text{s}^{-1}$)}, which is subsonic and highly sub-Alfv\'{e}nic (discussed below), so that the field in the volume evolves quasi-statically as occurs in the corona. 


The flow is ramped up smoothly, held constant for a time, and then reduced to zero before the onset of the jet in each simulation. {This is a numerically convenient way in which to form the filament channel and inject magnetic free energy into the corona. The flow is subsonic and subAlfv\'{e}nic, so the field evolution is quasi-static; however, for our coronal scalings it is still over an order of magnitude faster than typically observed surface flows on the Sun. Therefore, we reduced the flow speed to zero well before jet onset in each simulation, to avoid any direct driving of the jet by the imposed flow. This separates completely the artificially fast energy-injection process from the dynamically self-consistent mini-filament eruption and reconnection-driven jet onset that occur later.} The length of the constant driving phase was varied between each run, such that for $\theta=+22^\circ, 0^\circ,$ and $-22^\circ$ the total driving time was $t_{d} = 300$ {(20 min)}, $350$ {(23 min 20 s)}, and $450$ {(30 min)}, respectively. Different driving periods were required as a result of the high sensitivity of the breakout reconnection to the background-field inclination angle. This is discussed further in \S \ref{subsec:breakout}. Each simulation was halted before the jet disturbance reached the top boundary of the domain. 

We use the Adaptively Refined Magnetohydrodynamics Solver \citep[ARMS;][]{DeVore2008} to solve the ideal MHD equations in the form
\begin{gather}
\frac{\partial \rho}{\partial t} + \boldsymbol{\nabla}\cdot(\rho \mathbf{v}) = 0,\\
\frac{\partial (\rho \mathbf{v})}{\partial t}+\boldsymbol{\nabla}\cdot(\rho \mathbf{v}\mathbf{v}) + \boldsymbol{\nabla} P -\frac{1}{\mu_{0}}(\boldsymbol{\nabla}\times\mathbf{B})\times\mathbf{B} = 0,\\
\frac{\partial U}{\partial t}+\boldsymbol{\nabla}\cdot (U\mathbf{v})+P\boldsymbol{\nabla}\cdot\mathbf{v} = 0,\\
\frac{\partial \mathbf{B}}{\partial t}-\boldsymbol{\nabla}\times (\mathbf{v}\times\mathbf{B})=0,
\end{gather}
where $t$ is the time, $\rho$ is the mass density, $P = \rho R T$ is the thermal pressure, $U = P/(\gamma-1)$ is the internal energy density, $\mu_{0}=4\pi$ is the magnetic permeability, and $\mathbf{B}$ and $\mathbf{v}$ are the 3D magnetic and velocity fields. An ideal gas is assumed with ratio of specific heats $\gamma = 5/3$. We use a uniform plasma density, temperature and pressure of $1.0$, $1.0$ and $0.01$ respectively, so the non-dimensional gas constant is $R=0.01$. {With coronal scalings the density and temperature are $4\times 10^{-16} \,\text{g}\,\text{cm}^{-3}$ and $1.2\times 10^{6}$ K, respectively.} The corresponding plasma $\beta$ $\approx 0.22$ in the background field and drops to {$\beta \approx 8.8\times 10^{-4}$} at the surface within the parasitic polarity. The initially uniform sound speed $v_{s} \approx 0.13$ {($130 \text{ km}\, \text{s}^{-1}$)}, whilst the Alfv\'{e}n speed varies from $v_{a} \approx 0.3$ {($300 \text{ km}\,\text{s}^{-1}$)} in the background field to $v_{a} \approx 4.85$ {($4850 \text{ km}\,\text{s}^{-1}$)} in the parasitic polarity. We used box sizes of $[0,120]\times[-20,100]\times[-20,20]$, $[0,160]\times[-20,20]\times[-20,20]$, and $[0,160]\times[-100,20]\times[-20,20]$ for the simulations with $\theta = +22^\circ, 0^\circ,$ and $-22^\circ$, respectively. Open, zero-gradient boundary conditions were used on the top and side boundaries, whereas the bottom is closed and line-tied with zero tangential velocity everywhere except in the region of boundary driving. Reconnection occurs through numerical diffusion in the simulations. As we are primarily interested in the flow dynamics and the evolution of the magnetic field, we neglected gravity, the associated density and temperature stratification, and the thermodynamic effects of thermal conduction and radiative losses on the plasma. These simplifications mean that we cannot make meaningful predictions of the plasma radiation signatures in our simulated events; however, we expect that they would have little consequence for the magnetic- and flow-field evolution in our low-$\beta$ system. 

\begin{figure}
\centering
\includegraphics[width=0.45\textwidth]{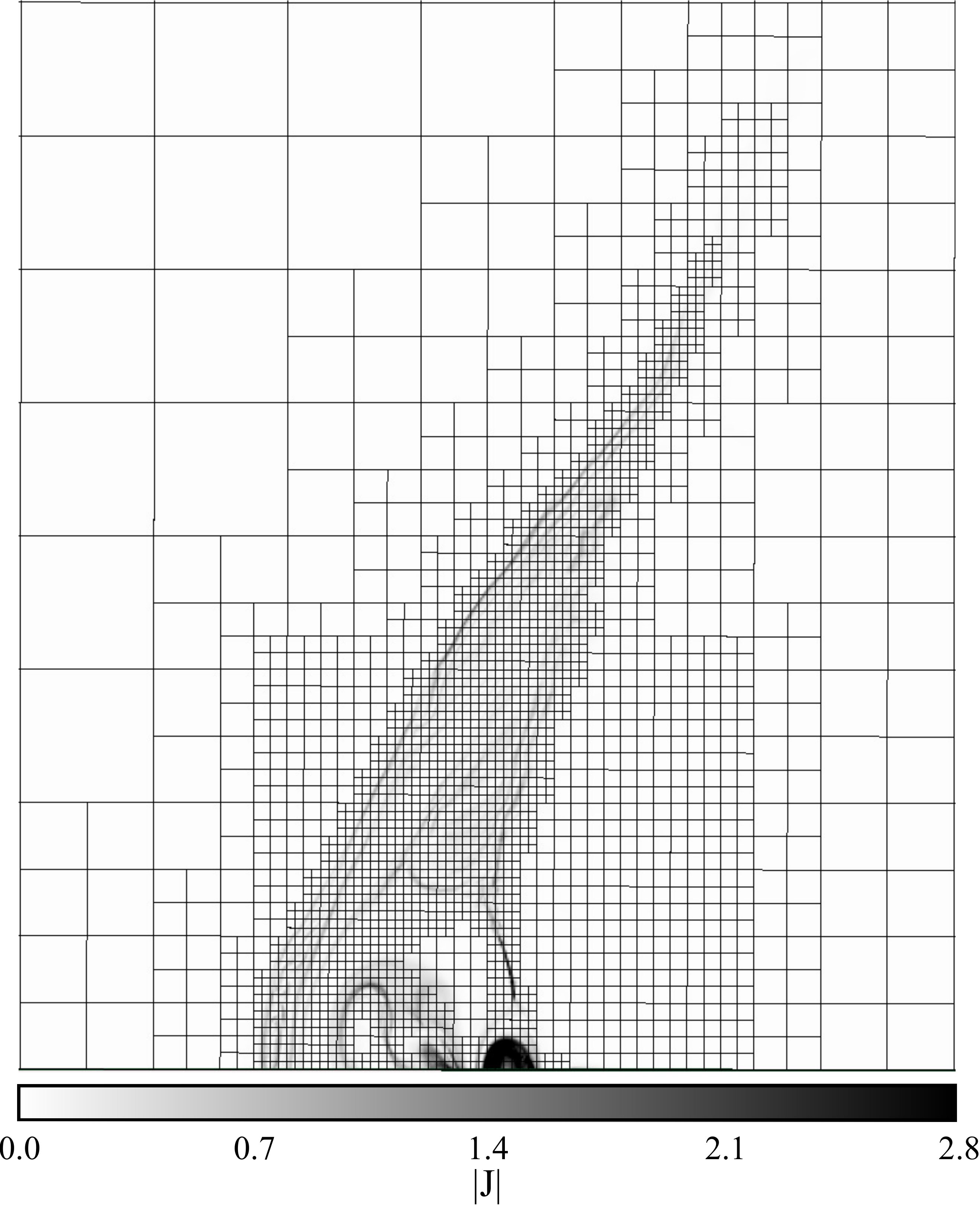}
\caption{The adaptively refined grid during the jet at $t = 31$ min $20$ s in the $\theta = +22^\circ$ simulation. Shown are cross sections of the grid blocks in the $z=0$ plane. Each block contains $8\times8\times8$ cells. The shading corresponds to non-dimensional electric current density ($\times \,1.5\times10^{-3} \text{A} \,\text{m}^{-2}$ with coronal scalings).}
\label{fig:grid}
\end{figure}

\begin{figure*}
\centering
\includegraphics[width=0.95\textwidth]{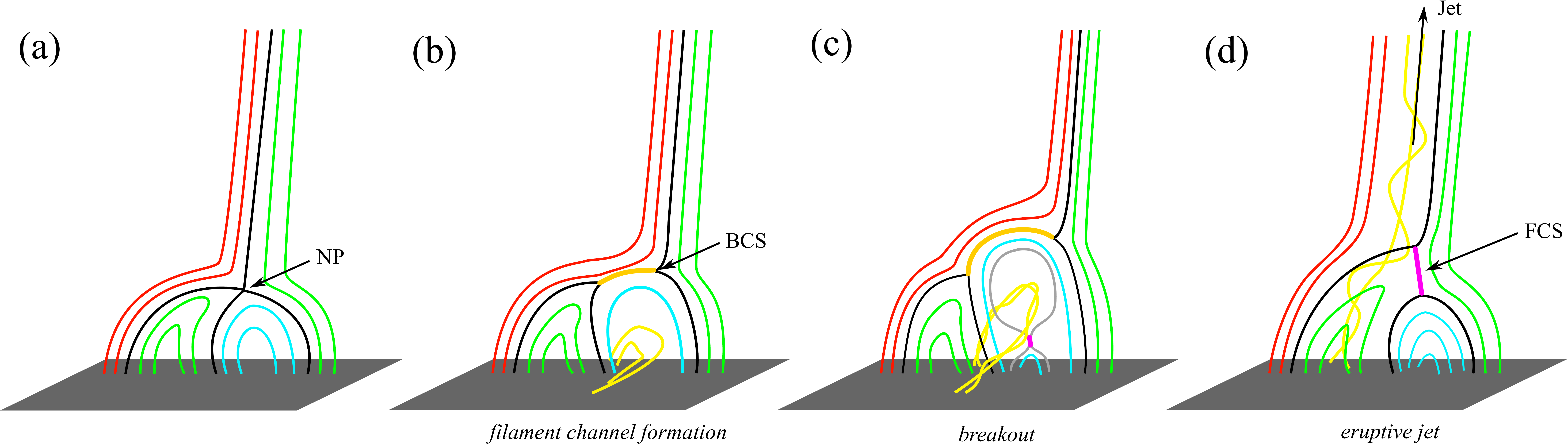}
\caption{Schematic of the evolutionary sequence that produces breakout jets. Green field lines show both the open (right) and closed (left) side-lobe regions. Cyan field lines below the null point show the strapping field that holds down the yellow field lines of the sheared filament/flux rope. Red field lines show the overlying background (open) field. Black field lines show the separatrix and spines of the null point. Grey lines show the cross section of the quasi-separatrix layer (hyperbolic flux tube) around (below) the flux rope. NP = null point, BCS = breakout current sheet, FCS = flare current sheet.}
\label{fig:fields}
\end{figure*}


The adaptive mesh employed by ARMS is managed using the PARAMESH toolkit \citep{MacNeice2000}. The grid refines/de-refines according to local measures of the gradient and field strength of the magnetic field \citep[see the appendix of][]{Karpen2012}. Extra resolution is added where there are high gradients in the magnetic field, such as at current sheets and shocks, and just as importantly is removed in regions lacking such features. Figure \ref{fig:grid} shows the block-adapted grid in the simulation with $\theta = +22^\circ$ during the evolution of the jet. Fine-scale grid blocks (each containing $8\times8\times8$ cells) outline both the jet front and the current layers on the separatrix dome. In each simulation, the background grid consists of blocks at the one level below the minimum refinement level shown in Figure \ref{fig:grid}. The grid was allowed to refine dynamically up to $5$ levels beyond this background. We imposed refinement to $4$ levels above the background in a small region that completely envelopes the separatrix dome, and to the maximum of $5$ levels in a thin layer that extends over the driving region at the surface. The adaptive-mesh capability of ARMS was crucial to resolving simultaneously the dynamics of the small-scale separatrix dome and those of the large-scale jet front.

\section{Jet Mechanism}
\label{sec:mechanism}
In each simulation, our system follows the same basic evolutionary sequence and exhibits four main phases: \textit{filament-channel formation}, \textit{breakout}, \textit{eruptive jet}, and \textit{relaxation}. A schematic of the first three phases of this sequence is shown in Figure \ref{fig:fields}. The configuration consists of just two distinct flux systems, open and closed field, separated by a null point (NP; Fig.\ \ref{fig:fields}a). Because the bipolar surface flux distribution is elongated in the out-of-plane direction of the figure, the evolution can be understood most easily by referring to the four-flux color scheme used in the figure: an internal closed-flux region that eventually hosts the filament (cyan field lines) is flanked by side lobes of both closed and open flux (green field lines) and is topped by oppositely-directed open flux (red field lines) on the far side of the null. This setup is topologically identical to the configuration investigated by \citet{Lynch2008}, who showed that it can give rise to large-scale breakout CMEs with eruptive flares. The main, but significant, difference is that in the setup for breakout CMEs studied by Lynch et al., the scale of the null separatrix is such that the ambient field strength declined with height above the photosphere. In addition, the external open-field region closed remotely back to the Sun. The role of the background field strength in suppressing or allowing ideal expansion during the evolution is the crucial factor that dictates the nature of the eruption (jet vs.\ CME) in the two setups \citep{Wyper2017}.

As the footpoint driving shown in Figure \ref{fig:setup}d proceeds, the large shear flow near the centre of the bipole forms a strongly magnetically sheared filament channel (yellow field lines) along and above the PIL (Fig.\ \ref{fig:fields}b). The rising magnetic pressure within the closed-field region expands the filament channel preferentially. Due to the strong strapping field (cyan field lines) overhead, the expanding sheared field increasingly stretches out to develop a quite flat midsection above the PIL \citep[see also][]{Antiochos1994,DeVore2000,Aulanier2002,DeVore2008}. Using 1D models with comprehensive descriptions of the thermodynamics, \citet{Karpen2001,Karpen2005} have shown that such regions can host long-lived condensations that resemble cool, counterstreaming filament plasma \citep[see also][]{Luna2012}. Over time, the null point above the strapping field becomes increasingly compressed, and a breakout current sheet (BCS; Fig.\ \ref{fig:fields}b) forms there.

Eventually, reconnection sets in at this sheet, removing some of the strapping field above the filament channel by transferring flux to the closed field under the far side of the dome and to the open field exterior to the near side of the dome (green field lines; Fig.\ \ref{fig:fields}c). The resultant upward lifting of the sheared field forms an initially weak current sheet (pink bar) below the filament. There are no null points within the channel, due to the strong out-of-plane field component. We infer from the presence of this current layer and the flux rope field lines that quasi-separatrix layers \citep{Titov2007} form around the filament and cross over beneath it (grey lines, Fig.\ \ref{fig:fields}c) at a hyperbolic flux tube \citep{Titov2002}. Slow reconnection occurs at their intersection, the weak current sheet, forming field lines that coil around the underside of the pre-existing filament. The growing flux rope rises at a slowly increasing rate determined principally by the removal of strapping field at the overlying breakout current sheet. At some point, the positive feedback between the removal of the strapping field and the rise of the flux rope reaches a critical threshold, beyond which eruption is inevitable \citep{Antiochos1999,Karpen2012,Wyper2017}.

Upon reaching the breakout sheet, the flux rope begins to reconnect rapidly with the external open field. This launches nonlinear Alfv\'{e}n waves that convect magnetic energy and compressed, accelerated plasma outward along the open field lines as the body of the jet (Fig.\ \ref{fig:fields}d). In addition, this rapid opening of the flux rope induces explosive interchange reconnection within the flare current sheet (FCS) left in its wake, producing the jet bright point. Subsequently, after the jet front has propagated away along the open field, the flare reconnection subsides and the closed-field region relaxes toward a new equilibrium configuration resembling the potential field with which we started (Fig.\ \ref{fig:fields}a).

\begin{figure}
\centering
\includegraphics[width=0.5\textwidth]{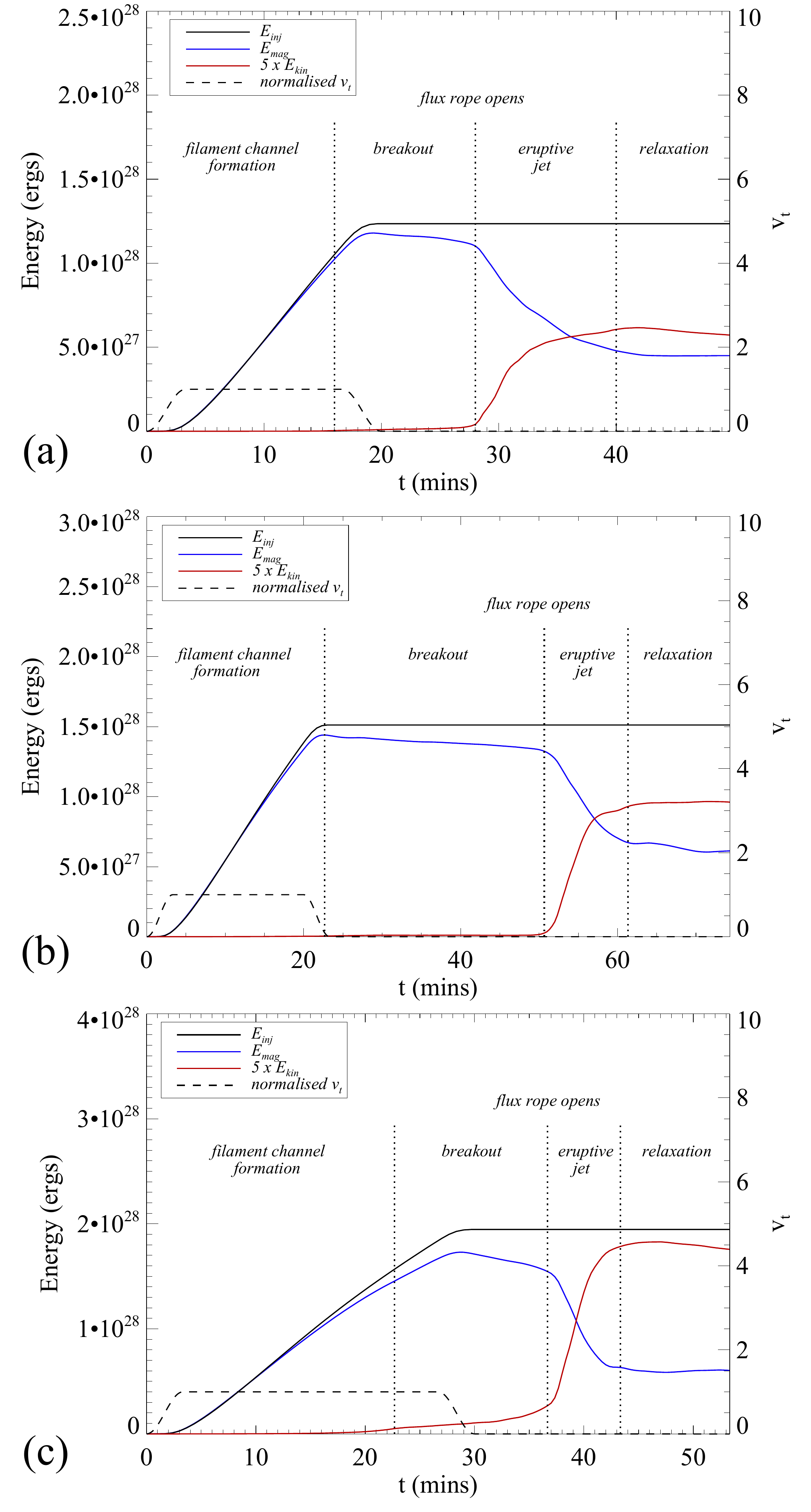}
\caption{Changes in free magnetic ($E_{mag}$, blue), kinetic ($E_{kin}$, red), and injected ($E_{inj}$, solid black) energies in each simulation. (a) $\theta = +22^\circ$. (b) $\theta = 0^\circ$. (c) $\theta = -22^\circ$. Dashed lines show the time dependence of the footpoint driving profile ($v_{t}$, normalised to unity). Note that different axes are used for energy and time in each plot and $E_{kin}$ is multiplied by $5$ for easier comparison.}
\label{fig:energies}
\end{figure}

\begin{figure*}
\centering
\includegraphics[width=0.9\textwidth]{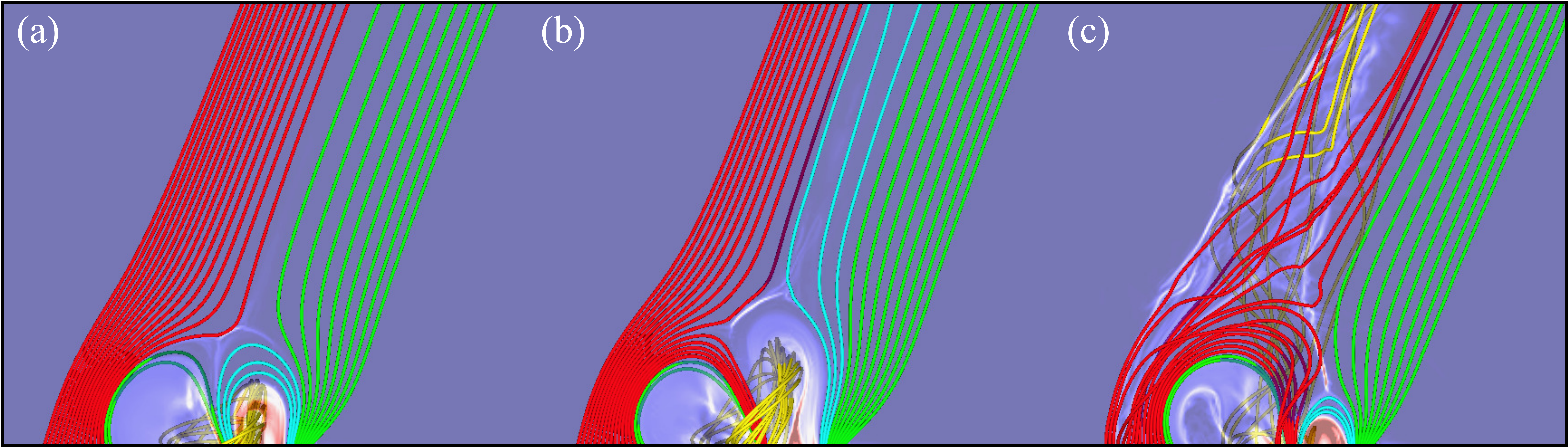}
\caption{The eruption sequence when $\theta = +22^\circ$. (a) $t = 16$ min, (b) $24$ min, and (c) $31$ min $20$ s. Shading shows electric current density ($|J|$) with the same colour scale as Figure \ref{fig:relax}. Red, cyan, and green field lines are traced from fixed, non-driven footpoints along the $y$-axis ($z=0$) on the photosphere. Yellow field lines that pass through the flux rope are traced from non-driven photospheric footpoints. An animation is available online.}
\label{fig:evolution1}
\end{figure*}

\begin{figure*}
\centering
\includegraphics[width=0.9\textwidth]{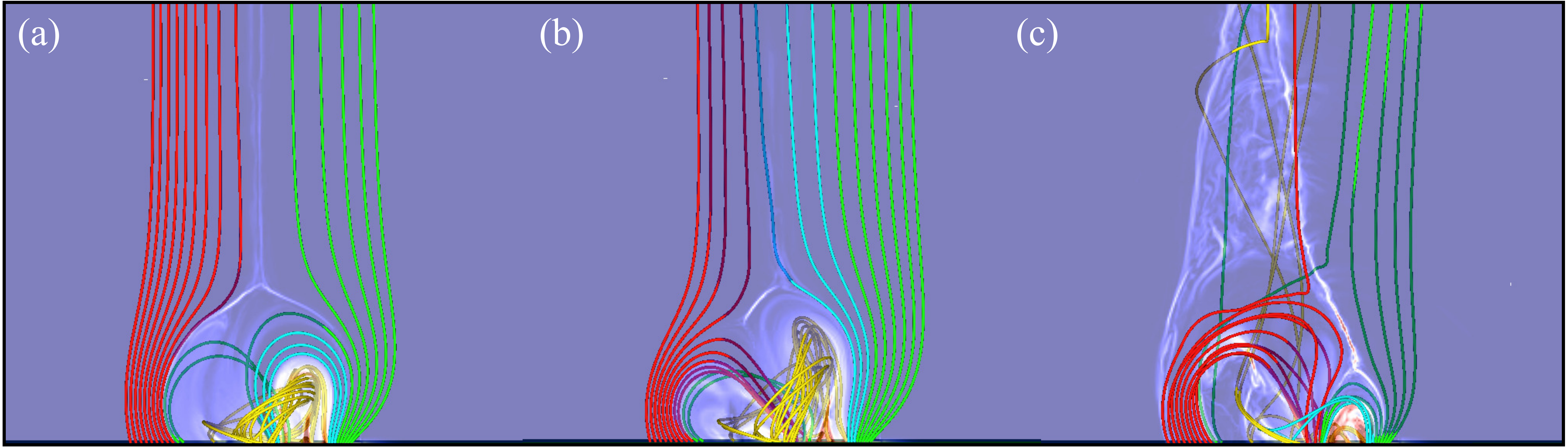}
\caption{The eruption sequence when $\theta = 0^\circ$. (a) $t = 22$ min $40$ s, (b) $45$ min $20$ s, and (c) $54$ min $40$ s. Shading and field lines as in Fig. \ref{fig:evolution1}. An animation is available online.}
\label{fig:evolution2}
\end{figure*}

\begin{figure*}
\centering
\includegraphics[width=0.9\textwidth]{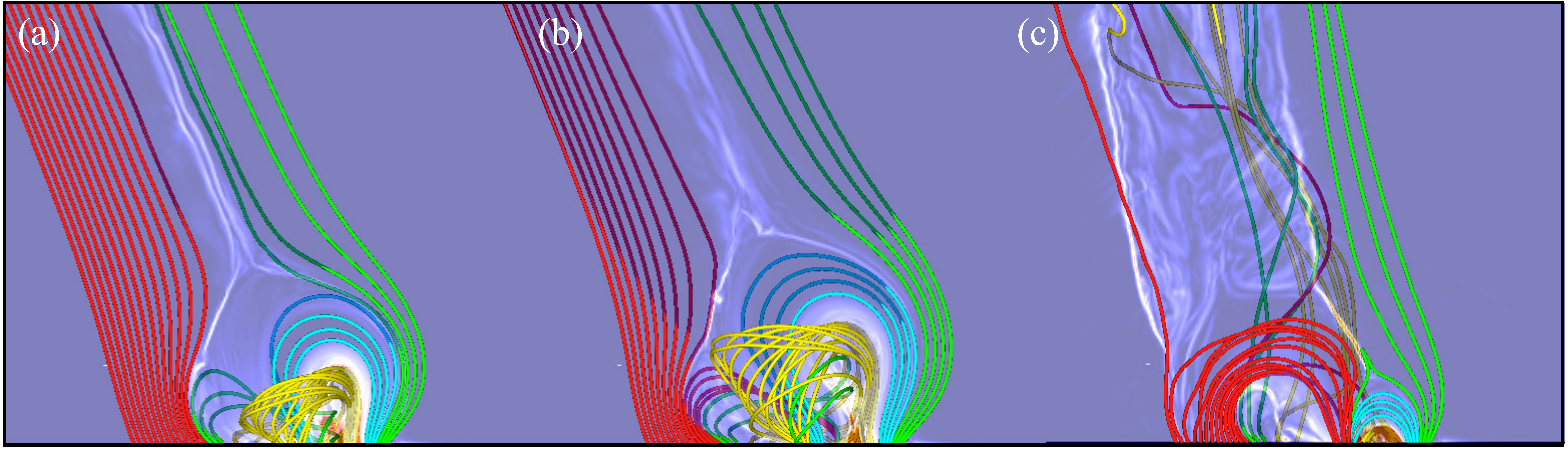}
\caption{The eruption sequence when $\theta = -22^\circ$. (a) $t = 23$ min $20$ s, (b) $30$ min $40$ s, and (c) $40$ min. Shading and field lines as in Fig. \ref{fig:evolution1}. An animation is available online.}
\label{fig:evolution3}
\end{figure*}

\begin{figure*}
\centering
\includegraphics[width=0.95\textwidth]{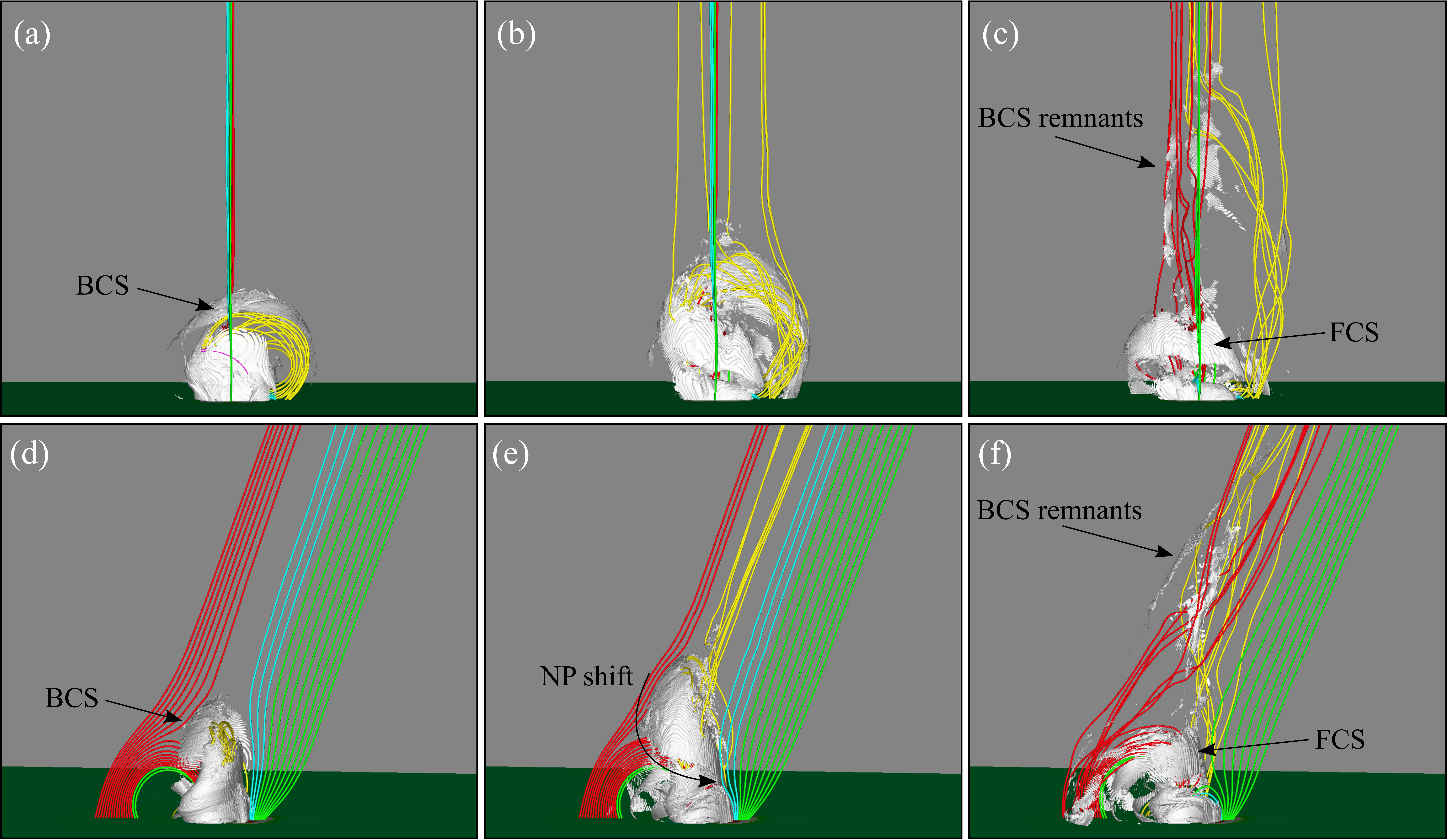}
\caption{End-on (top row) and side (bottom row) views of the shift of the interchanging current layer as the flux rope opens for $\theta = +22^\circ$. White iso-surfaces show $|J| = 1.0$ ($\times \,1.5\times10^{-3} \text{A} \,\text{m}^{-2}$ with coronal scalings). Field lines as in Figure \ref{fig:evolution1}.  Left column: $t = 27$ min $20$ s. Middle column: $t = 29$ min $20$ s. Right column: $t = 31$ min $20$ s. BCS = breakout current sheet. FCS = flare current sheet. NP  = null point.}
\label{fig:opening}
\end{figure*}

\begin{figure*}
\centering
\includegraphics[width=0.95\textwidth]{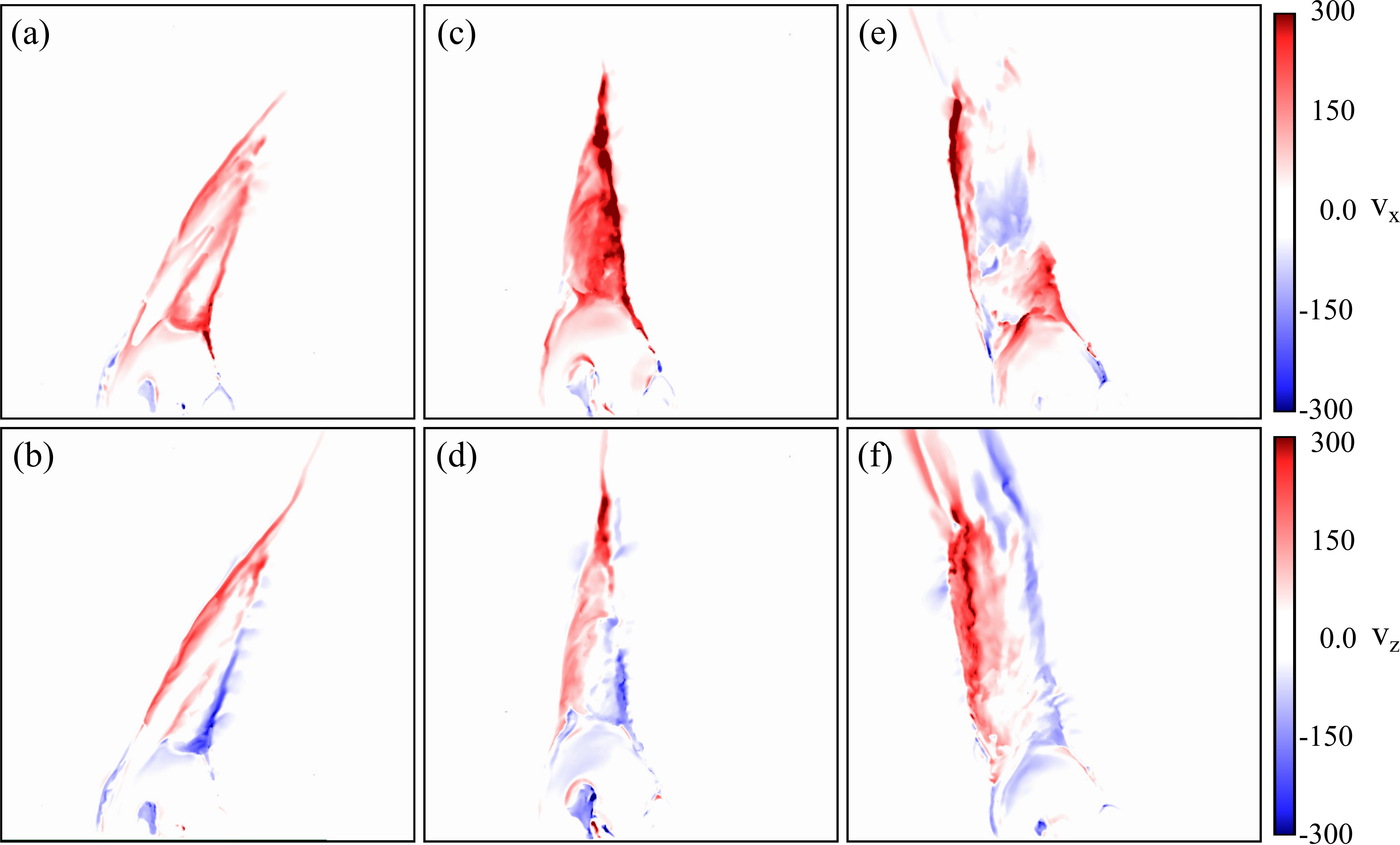}
\caption{$v_{x}$ (top) and $v_{z}$ (bottom) in the $z=0$ plane as the jet is launched. Left column: $\theta = +22^\circ$, $t = 31$ min $20$ s. Middle column: $\theta = 0^\circ$, $t = 54$ min $20$ s. Right column: $\theta = -22^\circ$, $t = 40$ min. All velocities are in km$\,\text{s}^{-1}$.}
\label{fig:velcuts}
\end{figure*}

\section{Overview of Energies and Morphologies}
\label{sec:overview}
The durations and onset times (vertical dotted lines) of the four evolutionary phases are indicated in Figure \ref{fig:energies}, where they are displayed along with the total free magnetic, kinetic, and injected energies. We calculated the energies from 
\begin{eqnarray}
E_{mag} &=& \iiint_V \frac{B^{2}}{8\pi} \,dV - \left[ \iiint_V \frac{B^{2}}{8\pi} \,dV \right]_{t=0}, \label{emag}\\
E_{kin} &=& \iiint_{V} \frac{1}{2}\rho v^2 \,dV, \label{ekin}\\
E_{inj} &=& \int{\left[\iint_{x=0} \frac{1}{4\pi} (\mathbf{v} \cdot \mathbf{B}) B_{x} \,dS \right]dt}. \label{einj}
\end{eqnarray}
Because the driving profile maintains the $B_{x}$ distribution on the photosphere and the simulations were halted before the jet reached the top or side boundaries, the lowest-energy (potential) magnetic field remains the same throughout the evolution in each case. Thus, $E_{mag}$ in Equation (\ref{emag}) represents the free energy stored within the magnetic field. $E_{kin}$ is the equivalent for kinetic energy, since the plasma is initially at rest. $E_{inj}$ is the cumulative injected Poynting flux across the photosphere due to the boundary driving. Under an ideal, quasi-static evolution, $E_{mag}$ should always equal $E_{inj}$ (neglecting small effects due to plasma energy).

There are significant differences between durations of each phase among our simulations, but the qualitative changes in the energies are quite similar in all cases. The filament-channel formation phase commences immediately, at $t = 0$, when the footpoint driving was turned on smoothly over a short ramp-up interval. Thereafter, the footpoint motion was held steady for some time, then turned off smoothly over a short ramp-down interval. Through experimentation, we found durations of steady motion that were sufficient in each case to generate an eruptive jet. The resulting driving profiles are the dashed curves shown in the figure. During this formation phase, $E_{kin}$ is negligible and $E_{inj}$ and $E_{mag}$ follow each other closely. This indicates that the evolution remains quasi-static and quasi-ideal, and nearly all of the energy injected by the boundary driving is stored as free magnetic energy.

{Some early reconnection and energy release occurs in the simulations with $\theta = 0^\circ$ and $-22^\circ$ towards the end of this phase. In these cases the footpoint of the inner spine of the null falls within the patch of surface motions, and so is displaced by the driving. This forms a current layer at the null point and drives reconnection that acts to add flux above the filament channel, further stabilising it. Both simulations were deemed to have transitioned to the breakout phase when the expansion of the filament channel overcomes this initial reconnection so that flux from above the filament starts to be removed in the manner of Fig. \ref{fig:fields}(b)-(c). For $\theta = +22^\circ$ the inner spine is undriven and the breakout phase begins once the breakout current sheet forms at the null and reconnection begins. }

Figures \ref{fig:evolution1}, \ref{fig:evolution2} and \ref{fig:evolution3} show a side view of each of our configurations at three times (left to right) during their evolution. The left column shows each case at approximately the time of transition from the filament-channel formation phase to the breakout phase. Field lines are traced from non-driven line-tied footpoints on the photosphere and are coloured the same as in our schematic diagram (Fig.\ \ref{fig:fields}). Colour shading in the $z=0$ plane shows current-density magnitude. The fan planes have elongated upwards substantially from the initial configurations shown in Figure \ref{fig:setup} and are outlined by moderately strong currents (white shading). The filament channels host the strongest electric currents (red shading) and strongly sheared magnetic fields (yellow field lines). The breakout current sheets, above and left of the filament-channel arcades of loops at the Y points of the external open field, have locally enhanced current densities.

As the breakout progresses, the filament-channel fields slowly distend upward towards the breakout current layer. This is evident in the middle column of Figures \ref{fig:evolution1}, \ref{fig:evolution2} and \ref{fig:evolution3}, whose images are taken from about halfway through this phase. One key feature to note here is the change in the innermost flux from the left open regions (red field lines): it has reconnected so that it now closes back to the Sun adjacent to the filament channel. This marks the progression of the breakout reconnection during this phase. The energy plot, Figure \ref{fig:energies}, shows that there is a gradually increasing deviation between $E_{inj}$ and $E_{mag}$ due to the quasi-steady release of stored magnetic energy by this reconnection. The release is slow, as evidenced by the very small to negligible $E_{kin}$ during this phase (notice that $E_{kin}$ is multiplied by 5 to improve its visibility in the figure).

The transition from the breakout to the eruptive-jet phase occurs when the rising flux rope in the filament channel began to reconnect with the external open field across the breakout current sheet. This was determined by examining the field lines threading the flux rope. This transition is an inherently three-dimensional and very dynamic process. It is shown in Figure \ref{fig:opening} for $\theta = + 22^\circ$. The left panels show the current structures just prior to the flux rope opening. The breakout current sheet curves over the top of the rising flux rope, shown as a curved iso-surface of $J$. Beneath this a strong volumetric current outlines the shape of the flux rope and includes the current layer below the rope. The middle panels show a time soon afterwards where the flux rope is beginning to open (yellow field lines). At this time, the breakout current sheet combines with the current layer beneath the flux rope, forming an extended current structure that wraps around the separatrix surface. As this occurs, the interchange reconnection region (effectively the null point, or cluster of null points, within the current structure) moves through the curved current structure from the top of the dome to behind the opening flux rope. This region becomes the explosively interchanging flare current sheet. The right panels show the flare current sheet once it is fully formed. The remnants of the breakout current sheet now form the filamentary current layers that separate the untwisting flux rope from the ambient field and propagate away with the jet.

The magnetic-field and current-density structures at about the mid-point of the eruptive-jet phase are shown in the right columns of Figures \ref{fig:evolution1}, \ref{fig:evolution2} and \ref{fig:evolution3}. The initially fully open (red) flux has now almost completely closed down to the surface as a consequence of the breakout reconnection. In each case, a broad spire of intense, filamentary currents extends upward into the corona from the reconnected-flux region, bordered on the left of the image by a strip of strong current demarcating the remnants of the breakout current sheet. These currents are markers of the nonlinear Alfv\'{e}n waves launched onto open field lines. Figure \ref{fig:velcuts} shows the vertical ($v_x$; top row) and out-of-plane horizontal ($v_z$; bottom row) components of the supersonic plasma flow in the three cases (left to right). The vertical flow is generally outward over most of the jet volume. The horizontal flow, in contrast, reverses direction across the center of the jet body. This indicates a rotational or torsional motion of the plasma, as the magnetic twist transferred from closed to open field is carried away as an untwisting wave. The resulting helical motions closely resemble those observed in many solar jets.

As can be seen in Figure \ref{fig:energies}, concurrent with or very soon after this opening of the flux rope, there is a steep drop in $E_{mag}$ and a simultaneous sharp increase in $E_{kin}$. These changes mark the sudden onset of the sustained, explosive interchange reconnection. The launching of the non-linear Alfv\'{e}n waves, together with the plasma acceleration within the flare current sheet, converts between $25\%$ and $35\%$ of the released free magnetic energy to kinetic energy of bulk flow in the jet, amounting to some $10^{28}$ erg. The associated durations of the eruptive-jet phase range from 6 min to 12 min. These energies and durations are consistent with those observed in coronal jets \citep{Shibata1992,Savcheva2007}.

Late in the eruptive-jet phase, the jet front has propagated away from the separatrix dome and the explosive interchange reconnection slows. The previously rapid changes in $E_{mag}$ and $E_{kin}$ tail off, as can be seen in Figure \ref{fig:energies}. Each jet then enters a relaxation phase, during which both the reduced free magnetic energy and increased kinetic energy remain at nearly constant values. The high efficiency of the jet-generating reconnection processes is indicated by the fact that more than $50\%$ of the initially stored free magnetic energy has been liberated by the time of transition to the relaxation phase.

\begin{figure*}
\centering
\includegraphics[width=0.95\textwidth]{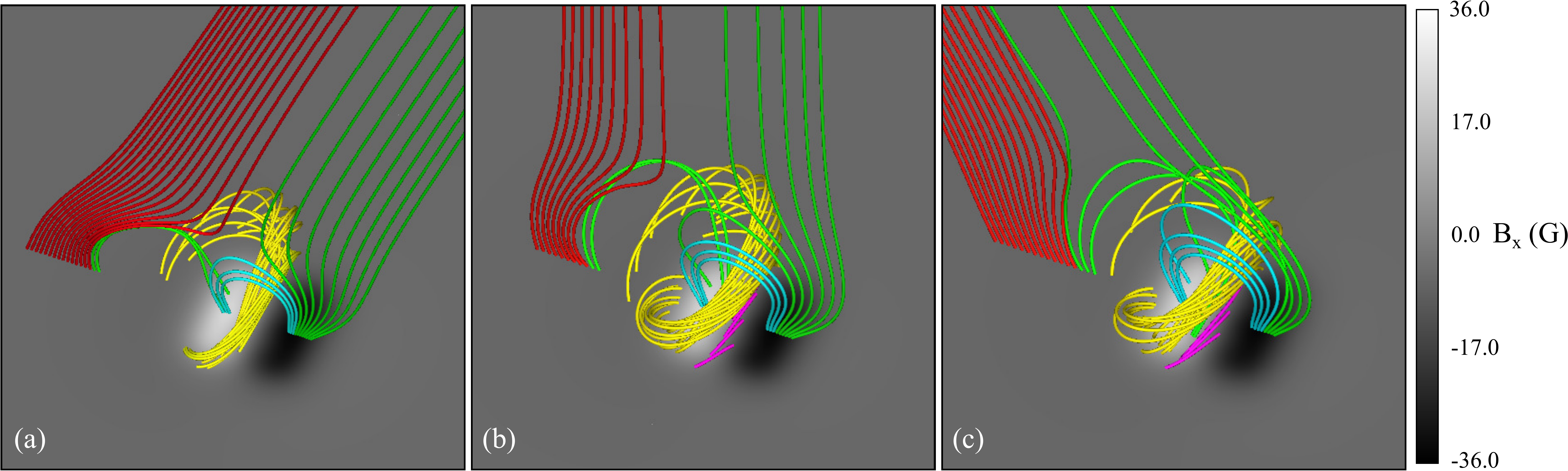}
\caption{Filament-channel field (yellow magnetic-field lines). Bottom plane is color-shaded according to $B_x$. (a) $\theta = +22^\circ$, $t = 16$ min. (b) $\theta = 0^\circ$, $t = 22$ min $40$ s. (c) $\theta = -22^\circ$, $t = 21$ min $20$s. In (b) and (c) shown in pink are the short, reduced shear field lines that form as the sheared arcade is converted to a flux rope by reconnection near the PIL.}
\label{fig:filaments}
\end{figure*}

\begin{figure*}
\centering
\includegraphics[width=0.95\textwidth]{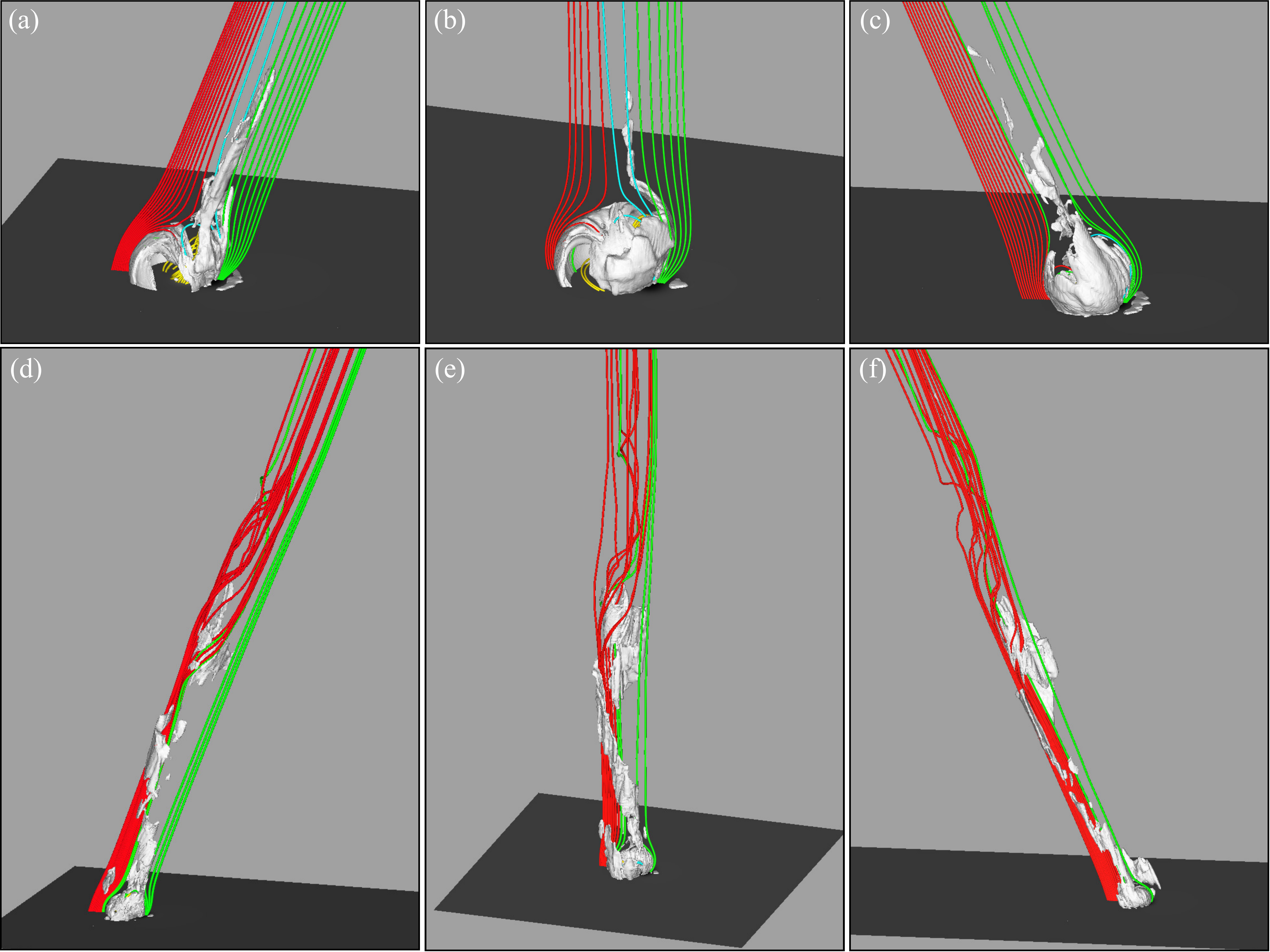}
\caption{Morphology of the jet in each simulation. Top row: during the breakout phase. Bottom row: during the eruptive-jet phase. Bottom plane is color-shaded according to $B_x$ as in Figure \ref{fig:filaments}. Isosurfaces show mass density $\rho = 1.1$ ($4.4\times 10^{-16} \,\text{g}\,\text{cm}^{-3}$). (a) and (d): $\theta = +22^\circ$, $t = 22$ min $40$ s and $t = 42$ min. (b) and (e): $\theta = 0^\circ$, $t = 40$ min $40$ s and $t = 61$ min $20$ s. (c) and (f): $\theta = -22^\circ$, $t = 24$ min and $t = 53$ min $20$ s. The separation of the leading fast nonlinear Alfv\'{e}n wave (indicated by the kinking of the open field lines) from the trailing slower plasma outflow (depicted by the density isosurfaces) is evident during each jet. Animations are available online.}
\label{fig:jetiso}
\end{figure*}

\section{{Detailed Comparison of Evolutionary Phases}}
\label{sec:phases}

\subsection{Filament-Channel Formation}
\label{subsec:filament}
{Since the filament forms deep within the closed field region and away from the influence of the background field inclination, the filament channel formation phase is similar in each simulation. However, when $\theta = -22^\circ$ and $0^\circ$ the onset of the breakout phase is delayed long enough (see below) that the highly sheared field within the filament channel is converted to a flux rope by so called ``tether-cutting" reconnection \citep{Moore1992,Moore2001}, Fig. \ref{fig:filaments}(b) and (c). This connectivity change is induced within a narrow vertical current layer that forms along the PIL as the field there is sheared. The reconnection lengthens the higher sheared field lines, whilst also adding twist to form a flux rope. Additionally, less sheared field lines are formed beneath the rope (Fig. \ref{fig:filaments}, pink field lines) \citep{vanBallegooijen1989}. This slow reconfiguration of the field releases negligible amounts of free energy (Fig. \ref{fig:energies}), and leads to no rapid dynamics. These flux ropes are actually remarkably robust, and in test simulations where the driving was halted after the flux ropes form, but prior to the onset of the breakout phase, the system would find a new equilibrium with the flux rope embedded in the closed field region. These results tell us three important things: (1) tether cutting reconnection is not the driver of these jets, (2) ideal instabilities of the flux rope, such as kink or torus, are also not the drivers, and (3) the subsequent breakout behaviour is not sensitive to whether a sheared arcade or true flux rope is present initially. }

\subsection{Breakout}
\label{subsec:breakout}
The inclination angle of the field plays an important role in the formation of the breakout current sheet and, subsequently, in the onset of the eruptive jet. For large positive values of $\theta$, the background field is in the opposite direction to the horizontal field of the compact bipole. Consequently, the null point is positioned more or less directly above the parasitic polarity of the bipole (Fig.\ \ref{fig:setup}a). The expanding strapping field pushing up into the oppositely directed background field then readily forms the breakout current sheet there. This explains both the early onset and the comparatively short duration of the breakout phase for the case $\theta = +22^\circ$ (Fig.\ \ref{fig:energies}, top).

As $\theta$ is reduced and the background-field orientation rotates to vertical and beyond, the null point moves farther from the PIL of the bipole. This shifts the photospheric footpoint of the inner spine of the null farther away from the PIL, and increases the amount of strapping field above the PIL where the filament channel will form (Fig.\ \ref{fig:setup}c). As the null is positioned farther to the side, the upward expansion of the strapping field above the filament channel less readily pushes into the null and the breakout current sheet is formed later. This is compounded in our simulations by the rotational driving profile that shears the field on both sides of the dome, so that the dome as a whole expands upwards. Thus, as the inclination angle changes, the effect on the null point changes from mainly compression across the fan plane (from pushing into the overlying field) with fast breakout-sheet formation to mainly stretching along the fan plane (from mis-matched expansion of the dome) with slow breakout-sheet formation. In our tests, we found that driving the configuration with $\theta = 0^\circ$ for the same duration as $\theta = +22^\circ$ was insufficient to initiate strong breakout reconnection; instead, the system reached a new equilibrium. Clearly, a critical threshold of breakout reconnection must be achieved to initiate a jet, just as occurs in CME calculations \citep[e.g.][]{Karpen2012}. The threshold would appear to be related to the balance of forces within the closed-field region. The breakout phase for $\theta = 0^\circ$ was almost twice as long as that for $\theta = +22^\circ$ (Fig.\ \ref{fig:energies}). This is primarily due to the additional strapping field that must reconnect across the breakout current sheet for $\theta = 0^\circ$.

One further consequence of the shift of the null position is that as $\theta$ is reduced and the footpoint of the inner spine migrates away from the PIL, it moves farther into the driving region. In our configurations, the inner spine was driven not at all for $\theta = +22^\circ$, only slightly for $\theta = 0^\circ$, but quite strongly for $\theta = -22^\circ$. Driving the spine concentrates the shear at the null, directly forming a near-singularity in the current \citep{Pontin2007,Wyper2014a} rather than the broad breakout sheet formed by the expansion of field from below. The resultant boundary-driven reconnection adds additional strapping field above the filament channel, and consequently has a further stabilising influence. This effect dominates the early stages of the $\theta = -22^\circ$ evolution, which together with the extra strapping field and positioning of the null away from the PIL required significantly more driving to initiate the breakout process (Fig.\ \ref{fig:energies}, bottom). Tests with shorter driving durations all reached new equilibria following an interval of reconnection at the null. By the time the breakout phase started, most of the flux beneath the dome had become strapping field above a large filament-channel region (Fig.\ \ref{fig:evolution3}a). Despite this, the ensuing breakout phase is shorter for $\theta = -22^\circ$ than for $\theta = 0^\circ$. This is a result of the intense breakout reconnection facilitated by the strong breakout current sheet in this case.

Figure \ref{fig:jetiso} (top row) shows iso-surfaces of mass density depicting the compressed exhaust plasma of the breakout current sheet in each jet. For $\theta = +22^\circ$, the outflows form a tapered spire (Fig.\ \ref{fig:jetiso}a) that waves and undulates (see the online movie). This wave motion follows the onset of tearing in the breakout sheet, in which blobs of high density plasma associated with small flux ropes are formed in and ejected from the sheet \citep{Wyper2014b,Wyper2016b}. Plasma in the spire is ejected at around $150 \,\text{km}\,\text{s}^{-1}$, only marginally above the background sound speed of $130 \,\text{km}\,\text{s}^{-1}$. The strongest, nearly Alfv\'{e}nic flows are concentrated downwards over the surface of the separatrix. For $\theta = -22^\circ$, the spire is less coherent and more fragmented (Fig.\ \ref{fig:jetiso}c). The outflows have a speed near the local Alfv\'{e}n speed, $\approx 300 \,\text{km}\,\text{s}^{-1}$. For $\theta = 0^\circ$, little density enhancement occurs as these outflows too are directed over the separatrix surface. A thin, transient spire is visible in Figure \ref{fig:jetiso}(b). These results suggest that regions of the corona with highly inclined open fields should exhibit outflows from the breakout reconnection that are visible as straight jet-like spires. In nearly vertical fields, on the other hand, the jet spire should be weak or even unobservable during the breakout phase.

\subsection{Eruptive Jet}
\label{subsec:jet}
In each of our configurations, a violent change in behaviour occurs when the flux rope reaches the breakout current sheet. Interchange reconnection opens the end of the flux rope previously rooted in the parasitic (positive) polarity of the bipolar region (yellow field lines: Figs\ \ref{fig:evolution1}, \ref{fig:evolution2} and \ref{fig:evolution3}, right panels). This launches a nonlinear torsional Alfv\'{e}n wave as part of the twist within the flux rope propagates outwards along the reconfigured open field lines in the manner first envisaged by \citet{Shibata1986}. Plasma around the periphery of the unwinding flux rope is driven upwards in a spiral with strong out-of-plane and upward components (Fig.\ \ref{fig:velcuts}). However, the untwisting wave is only one aspect of each jet. The shift of the interchange reconnection site (described in \S \ref{sec:overview}) initiates explosive flare reconnection in the current sheet behind the rope. The reconnection accelerates plasma upwards into the underside of the untwisting wave front and downwards into low-lying flare loops. This bi-directional outflow is clear on the right side of the top panels in Figure \ref{fig:velcuts}. Note that like the breakout current sheet, the flare current sheet is fragmented, which creates substructure in $v_{x}$ within the sheet. The sheet strengthens and reconnects explosively as the flux transferred from above the bipole to the other side of the dome slams back into the ambient field in the wake of the flux-rope ejection (red field lines in Figs\ \ref{fig:evolution1}, \ref{fig:evolution2} and \ref{fig:evolution3}, right panels). 

Figure \ref{fig:jetiso} (bottom row) shows the untwisting jets that are formed. The kinked field lines (predominantly red) show part of the torsional wave launched by the opening of the flux rope. Isosurfaces of density highlight the compressed plasma that forms part of the jet outflow. The kinked field lines and density enhancements appear together but gradually separate as the two propagate outward (seen clearly in the online movies), with the torsional wave front traveling at the local Alfv\'{e}n speed ($\approx 300 \,\text{km}\,\text{s}^{-1}$) and the density enhancement closer to the local sound speed ($\approx 130 \,\text{km}\,\text{s}^{-1}$). \citet{Pariat2016} recently described a very similar behaviour in their coronal-jet simulations, attributing the formation of the slower density enhancement to plasma accelerated by the passage of the Alfv\'{e}n wave. It seems likely that the same scenario is occurring here.  

The sharp decrease in magnetic energy and increase in kinetic energy vary across the three configurations (Fig.\ \ref{fig:energies}). For $\theta = +22^\circ$, the drop in magnetic energy begins steeply and progressively tails off as the jet proceeds over a duration of $\approx 12$ min. As $\theta$ increases, the energy-release interval shortens slightly to $\approx 10$ min $40$ s for $\theta = 0^\circ$ and then more dramatically to $\approx 6$ min $40$ s for $\theta = -22^\circ$. Correspondingly, the fraction of the free energy that is converted to kinetic energy increases from $\approx 25\%$ for $\theta = +22^\circ$ to $\approx 35\%$ for $\theta = -22^\circ$. The shortening of the jet period and increase in kinetic-energy conversion can be understood by considering where the magnetic energy is stored and released in each configuration. The free energy is injected as shear into the closed field, with the majority being found in the filament channel. For $\theta = 0^\circ$ and $-22^\circ$, weak tether-cutting reconnection creates a flux rope from the highly sheared field within the channel, transferring shear from low-lying field above the filament channel to the developing flux rope. This creates longer flux-rope field lines and shorter reconnected loops beneath. An increasing fraction of the free energy stored in the closed field resides within the flux rope. Once the flux rope begins to erupt, additional tether-cutting reconnection lengthens it so that it extends farther around the circular polarity inversion line and receives more of the free energy stored within the structure. Thus, the increasing duration of driving for the simulations with progressively smaller $\theta$ stores more free energy within the flux rope. The more impulsive energy release for smaller values of $\theta$ follows from a greater proportion of the stored free energy in the closed-field region being released promptly as the flux rope opens. The increased fraction of the free energy being converted to bulk kinetic energy then can be understood as resulting from a greater direct ideal acceleration of plasma by the untwisting torsional wave front. 

\begin{figure*}
\centering
\includegraphics[width=0.95\textwidth]{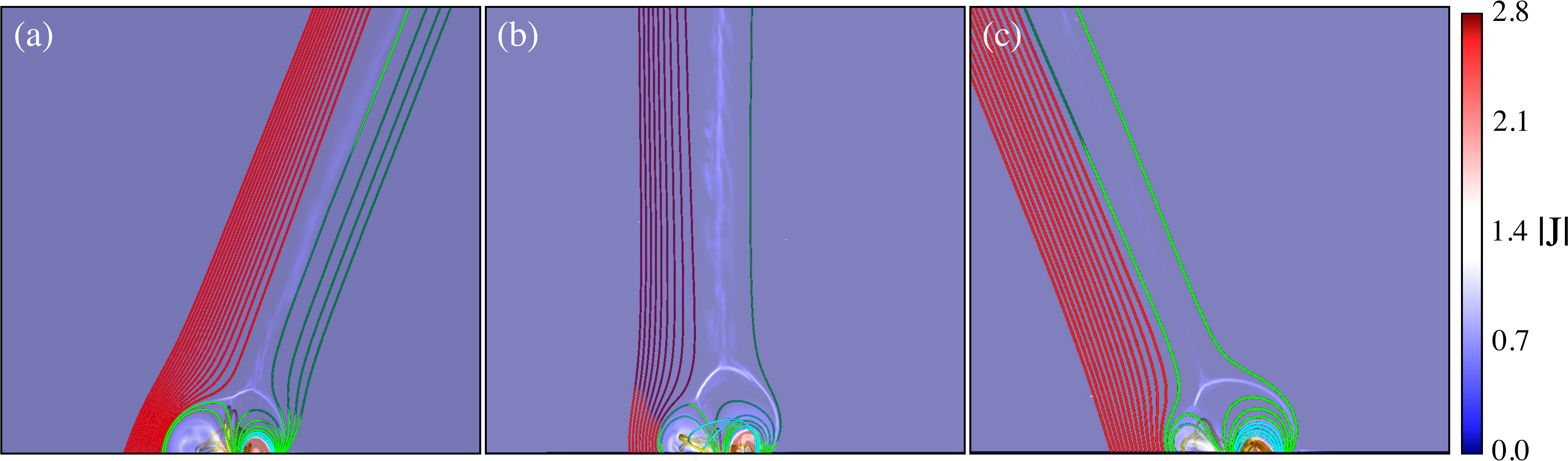}
\caption{The post-jet magnetic field. (a) $\theta = +22^\circ$, $t=49$ min $20$ s. (b) $\theta = 0^\circ$, $t= 66$ min $40$ s. (c) $\theta = -22^\circ$, $t=54$ min. Shading shows the non-dimensional electric current density $\vert {\bf J} \vert$ ($\times \,1.5\times10^{-3} \text{A} \,\text{m}^{-2}$ with coronal scalings).}
\label{fig:relax}
\end{figure*}

\begin{figure*}
\centering
\includegraphics[width=0.95\textwidth]{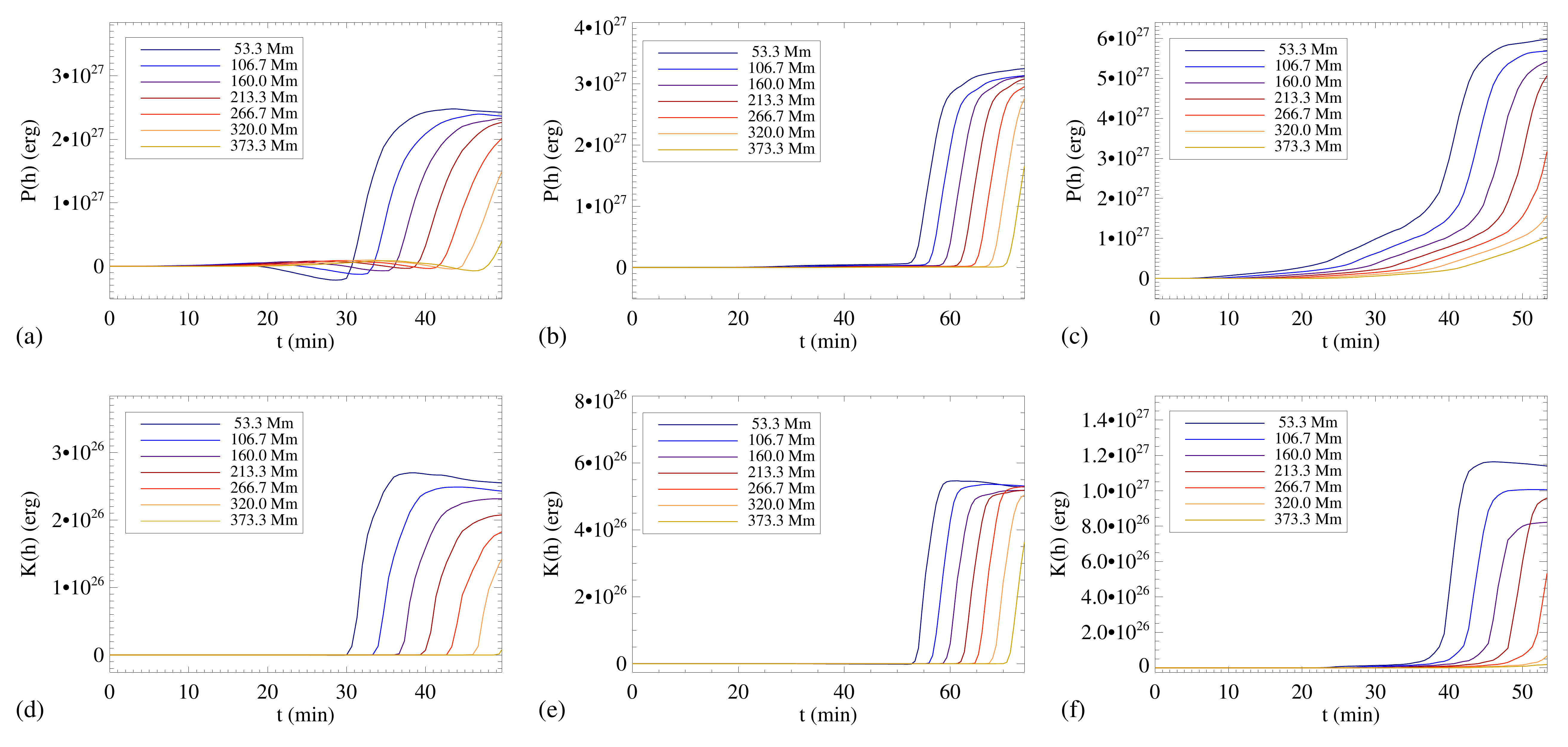}
\caption{Top: cumulative Poynting flux $P(h)$ across surfaces at selected heights $h$ above the photosphere. Bottom: cumulative kinetic-energy flux $K(h)$ across the same surfaces. (a) and (d): $\theta = +22^\circ$. (b) and (e): $\theta = 0^\circ$. (c) and (f): $\theta = -22^\circ$. Note the different vertical and horizontal scales used in each graph.}
\label{fig:fluxes}
\end{figure*}

\subsection{Relaxation}
Figure \ref{fig:relax} shows the field configurations near the bipole in the aftermath of the jets. In each case, the null dome resets to a configuration similar to the initial condition, but with a slowly reconnecting current layer at the null point. Some magnetic shear also remains in the closed-field region (particularly within the filament channel), as shown by the contours of strong volumetric current (white and red shading). Only a fraction of the shear on any given closed field line is released to propagate away when it is interchange-reconnected \citep[e.g.][]{Wyper2016b}. The remnant sheared field contributes the majority of the free magnetic energy at the end of the simulations (Fig.\ \ref{fig:energies}). The interchange reconnection at the null continues to progressively tail off as the closed-field region relaxes toward a new equilibrium. In each jet, the interchange reconnection continued until the simulation was halted so that the jet front did not reach the top boundary. The free energy released during this relaxation phase was negligible.

\section{Coronal Energy Injection}
\label{sec:fluxes}
Once the jet is launched, it propagates upwards along the ambient field, transporting energy (and also helicity) higher into the corona. To understand the details of this process, we calculated the cumulative energy transfer due to the Poynting and kinetic-energy fluxes across several different heights in each simulation. Specifically, we calculated 
\begin{eqnarray}
P(h) &=& \int{\left[\iint_{x\,=\,h} \frac{1}{4\pi}\left[ (\mathbf{v} \cdot \mathbf{B}) B_{x} - B^{2} v_{x} \right] dS \right]dt}, \label{pflux} \\
K(h) &=& \int{\left[\iint_{x\,=\,h} \frac{\rho v^{2}}{2} v_{x}  \,dS \right]dt}, \label{kflux}
\end{eqnarray}
where $P$ is the cumulative Poynting flux, $K$ the cumulative kinetic-energy flux, and $h$ is the height above the surface. The first term of $P$ is the contribution from motions tangential to the surface (the shear component), whilst the second corresponds to contributions from emergence/submergence of magnetic field (the vertical component). The results are shown in Figure \ref{fig:fluxes}. 

The cumulative Poynting flux during the breakout phase behaves differently in each case. For $\theta = +22^\circ$ the Poynting flux is negative, for $\theta = 0^\circ$ it is essentially zero, and for $\theta = -22^\circ$ it is positive. During the breakout phase, both $v_{x}$ and $v_{z}$ are small at all of these heights above the surface, and the dominant contribution to the Poynting flux is from the $v_{y}B_{y}$ term directed perpendicular to the PIL of the bipolar region. (In contrast, the dominant contribution at the surface $h=0$ is from the $v_{z}B_{z}$ shear term, which generates most of the injected energy $E_{inj}$ discussed in \S \ref{sec:overview}). As the breakout reconnection proceeds, the open field in each configuration moves in the positive $y$ direction (from left to right in Figs\ \ref{fig:evolution1}, \ref{fig:evolution2} and \ref{fig:evolution3}, for example) as it approaches the breakout current sheet, reconnects through the sheet, and then departs the sheet along with the reconnection exhaust. This produces a negative Poynting flux for $\theta = +22^\circ$ ($B_{y} < 0$), a negligible net flux for $\theta = 0^\circ$ ($B_{y}=0$), and a positive Poynting flux for $\theta = -22^\circ$ ($B_{y} > 0$). These results imply that the energy in the overlying magnetic field {directly above the bipole} decreases, remains about the same, and increases, respectively, in the three configurations. 

In all cases, once the jet is launched, a strong positive Poynting flux dominates as the torsional Alfv\'{e}n wave propagates upwards into the domain. The curves at greater heights rise progressively later due to the time required for the wave to propagate to those higher altitudes. These impulsive increases are much larger in amplitude than the quasi-steady changes that occurred during the breakout phase. 

In contrast to the Poynting flux, the kinetic-energy flux is always negligible during the breakout phase, then increases impulsively once the jet is launched. Unlike $P$, the curves for $K$ at progressively greater heights do not decrease monotonically with height. At $266.7$ Mm for $\theta = 0^\circ$ (Fig.\ \ref{fig:fluxes}e) and $213.3$ Mm for $\theta = -22^\circ$ (Fig.\ \ref{fig:fluxes}f), for example, the kinetic-energy fluxes are slightly larger than those at the next lower height. This suggests that further conversion of free magnetic energy to kinetic energy is occurring within the propagating jet front. Comparing the magnitudes of $K$ and $P$, it is clear that the energy transfer is dominated by the Poynting flux. Although there is a significant upward acceleration of plasma in each jet (Fig.\ \ref{fig:jetiso}), most of the kinetic energy resides within the rotational velocity component of the torsional Alfv\'{e}n wave. 

For completeness, we also calculated the cumulative enthalpy flux in each jet. Like the kinetic-energy flux, we found the enthalpy flux to be significantly smaller than the Poynting flux. Our results imply that the energy injected into the solar wind by coronal jets is far larger than what would be inferred from observations of only the jet plasma.  

\section{Discussion}
\label{sec:discussion}

\subsection{Comparison to Observations}
The jets produced by our model closely match several aspects of coronal jets involving mini-filaments. Jet mini-filaments are typically close in size to the width of the jet base, usually assumed to correspond to the closed-field region. Consequently, the lengths of mini-filaments vary with jet size, with quoted values varying from $l \approx 6$ Mm to $l \approx 36$ Mm \citep[e.g.][]{Sterling2015,Panesar2016}. In our simulations, the sheared filament channel is of comparable size to the extent of the separatrix in the $z$ direction. With our chosen scale values this is $l = 28$ Mm, falling within the range of observed values. Our filament channel also forms along the PIL of the strong pre-existing bipolar field, as is observed \citep[e.g.][]{Adams2014,Panesar2016}. Due to the highly simplified atmosphere and energy equation that we adopted, the cool, dense material associated with solar mini-filaments is not present in our simulations. However, the strongly sheared magnetic structure of the filament channel is replicated by our model.

The eruption sequence exhibited by our simulated jets matches well with numerous observations. For the three cases that we studied, we found that the intensity of the pre-jet breakout reconnection depends upon the inclination of the ambient magnetic field: faster, denser outflows result when the field is highly inclined, whereas weaker outflows with less density contrast occur when the field is vertical. Some examples of blowout jets preceded by a tapered, inverted-Y-shaped jet have been reported \citep[e.g.][]{Liu2011,Hong2016,Zhang2016b}. The perspective makes it difficult to discern the inclination of the ambient field in the jet described by \citet{Hong2016}. In the jets discussed by \citet{Liu2011} and \citet{Zhang2016b}, on the other hand, the ambient field is highly inclined, which is consistent with our findings. A clear example of a mini-filament jet in a nearly vertical field was {described by \citet{Moore2015} and revisited in \citet{Wyper2017}.} A weak spire connects to the edge of a sharp interface between closed and open field, consistent with a weak tapered outflow from a breakout current sheet (forming the sharp interface) as seen in our jet experiment with $\theta = 0^\circ$ (e.g., Fig.\ \ref{fig:evolution2}b).

The jets are produced in our model by a combination of an untwisting flux rope and plasma accelerated by the flare reconnection that occurs below. The broad, untwisting jet spire is consistent with many blowout jets. So too is the formation of flare loops by magnetic reconnection across our low-lying, vertical flare current sheet. With full plasma thermodynamics included, these loops should be heated by the reconnection process and, thus, correspond to the jet bright points formed beneath the erupting mini-filaments in the observations \citep[e.g.][]{Sterling2015}. Our jet plasma is a combination of ambient material within the flux rope and material that has been processed by the flare interchange reconnection. This is consistent with the observed multi-thermal nature of many mini-filament jets \citep[e.g.][]{Adams2014,Sterling2015}, which appear to be comprised of both cool filament plasma and hot coronal plasma from the reconnection region. Our post-jet relaxation phase, during which the flare reconnection tapers off while producing a continued stream of plasma in the wake of the main jet, also seems to be a common feature of mini-filament jets \citep[e.g.][]{Liu2011}.

Our simulations do not include all of the physics necessary to produce chromospheric/photospheric brightening (e.g., thermal conduction, radiation, and possibly non-thermal particles). Nevertheless, based on the magnetic-field evolution in our jet model, we can make informed conjectures about the expected photospheric signatures. Spreading flare ribbons, similar in nature to those in large-scale two-ribbon flares, can be expected to form at the base of the new loops formed by the flare reconnection as the jet is launched. In addition, the intense interchange reconnection initiated as the flux rope reaches the breakout current sheet should create brightening around the base of the separatrix surface. The location of this brightening will shift as the interchange reconnection changes the footprint of the separatrix surface. Depending upon the nearby distribution of flux, this brightening could be quasi-circular, as in some large-scale solar flares \citep[e.g.][]{Masson2012}, or take the form of discrete patches, if the separatrix field predominantly connects to discrete sources. \citet{Zhang2016b} observed both the spreading small-scale flare ribbons and a larger-scale, quasi-circular, enclosing ribbon as the jet was launched. \citet{Hong2016} described a mini-filament jet in which the separatrix brightening occurred across several nearby discrete patches associated with discrete photospheric flux regions. Therefore, our jet model is also qualitatively consistent with these observations. However, to understand the nature and timing of the photospheric brightening in our model requires a detailed analysis of the changing magnetic topology. This task is left to future work. 

Finally, as the jet front propagated outwards in our simulations, we observed a separation of the strong magnetic-field perturbation from the bulk plasma flow. The former propagated at the local Alfv\'{e}n speed, whilst the latter traveled at close to the local sound speed. Similar simulation results have been reported by \citet{Pariat2016}. Although we have not studied these features in any detail, we note that such a separation also has been reported for many observed jets \citep{Cirtain2007,Savcheva2007}. 

\subsection{Comparison to Previous Models}
From our numerical experiments, we identified the breakout-reconnection process as the dominant mechanism underlying the eruptive jet. In the picture presented, this reconnection initially is quasi-steady, creating a tapered outflow of plasma as envisaged originally by \citet{Shibata1992}. Eventually, the reconnection transitions to an explosive phase as the rising flux rope encounters the breakout current sheet and opens up. The resulting untwisting jet is similar to that conceived by \citet{Shibata1986}. There are some important differences between these early suggestions and our current work, however. First, our simulations produce both the quasi-steady tapered outflow and the subsequent impulsive jet in the absence of any flux emergence whatsoever. This suggests that these features of mini-filament jets, at least, may be universal and occur irrespective of whether flux is emerging within the jet source region. Second, there is positive feedback between the expansion of the filament-channel field below the breakout current sheet and the interchange reconnection of the strapping field across it. This is a key feature of the magnetic-breakout mechanism \citep{Antiochos1999} and provides the energy release needed to accelerate the explosive breakout reconnection process. Third, we find that the site of interchange (open/closed) reconnection moves from the breakout current sheet above the mini-filament flux rope to the initially slowly reconnecting current sheet below it, leading to the onset of explosive flare reconnection during the jet. This transition simultaneously launches the multi-thermal Alfv\'enic jet and produces the hot flare loops corresponding to the jet bright point.

The early Shibata models envisioned only a single current sheet, formed at the interface between the emerging (closed) and ambient (open) magnetic flux systems, so that the jet and the flare always would be in very close proximity to one another. In our model, the free magnetic energy is introduced by shearing the field along the PIL of the embedded bipolar region. The strong shear at the centre of the bipole induces reconnection near the photosphere that creates a flux rope in two of the configurations prior to the onset of the breakout phase. However, this weak tether-cutting reconnection \citep{Moore1992,Moore2001} neither significantly releases any of the stored energy nor initiates the eruption. More rapid tether-cutting reconnection occurred once the breakout phase began and the flux rope began to rise as strapping field was removed from above, as is expected in both the tether-cutting and breakout pictures. Thus, although tether cutting is certainly involved, and in fact is crucial for converting the shear in the filament-channel field to twist within the flux rope, it is not the driver of our jets.

This internal reconnection seems to play an important role in suppressing the global kinking of the closed field, which occurs in the kink-initiated jet models of Pariat and coworkers \citep[e.g.][]{Pariat2009}. We estimated the number of turns that could theoretically be achieved in our simulations by tracing field lines from the photosphere at a time prior to the onset of internal reconnection, assessing the highest number of turns at this point, and extrapolating it to the full duration of the footpoint driving. The number of turns were $N \approx 1.2$, $0.9$, and $0.7$ for $\theta = -22^\circ$, $0^\circ$, and $+22^\circ$, within the range of $N = 0.8$ to $1.4$ found by \citet{Pariat2010} to set off a kink. Nevertheless, we observed no global kinking. Nor was any obvious writhe or rotation of the flux rope observed, as is thought to trigger some large-scale filament eruptions \citep{Torok2005}. 

The ideal torus instability \citep{Kliem2006} is frequently cited as explaining the eruption of flux ropes in bipolar magnetic fields. The instability assumes a pre-existing flux rope, and occurs in bipolar fields where the strapping field strength drops off faster than a critical threshold. In the magnetic configuration for our simulations, the background field strength is uniform, negating this instability well away from our closed-field region. Within the closed-field region, the field strength does drop off towards the null, but then increases again beyond it. It is not clear whether the torus instability could operate in this configuration; an ideal treatment of the evolution would be necessary to be definitive. In any case, it is certain that the dynamics in our simulations are dominated by the non-ideal evolution of the breakout and flare current sheets, and these dynamics are at the heart of the eruptive-jet generation.

Finally, we note that some flux-emergence experiments \citep[e.g.][]{Archontis2013} have exhibited a similar evolution in the untwisting jets they produce as those from our model. Our breakout process relies rather generically upon the storage of free energy within the magnetic topology of a null point above a strong bipolar field. In the observations and in our model, the null point and bipole are pre-existing. However, the same topology can be created dynamically by flux emergence, as occurs in the Archontis \& Hood numerical experiments. Indeed, even large-scale breakout CMEs can be realised in flux-emergence experiments when the overlying field is correctly aligned \citep[e.g.][]{Archontis2008,Hood2012,Leake2014}. Thus, our model provides a rather general framework for interpreting events where free energy is stored along the PIL of a bipole in a null point topology. In principle, this storage could occur due to flux emergence, surface motions, or even flux cancellation. Once the free energy is stored there, the subsequent breakout behaviour is expected to be more or less the same. 


\subsection{Summary}
\label{subsec:summary}

In this work we have described in detail a new model for coronal jets involving mini-filament eruptions. Our model extends the well-known breakout model for large-scale CMEs \citep{Antiochos1998,Antiochos1999} to these much smaller events \citep{Wyper2017} and explains a number of their observed features. 

In our model, free energy is stored in a filament channel along the PIL of a pre-existing bipole. The sheared filament channel erupts as the jet is launched via the breakout process. Following on from the initial study of \citet{Wyper2017},  we studied three realisations of the model with varying background-field inclinations and found the breakout mechanism to work robustly in each case. In configurations where the field is highly inclined to the vertical, the breakout reconnection produces an inverted-Y-type reconnection outflow, similar in nature to outflows observed prior to mini-filament jets in similar configurations. This outflow was much weaker when the field is vertical. In all configurations, a broad untwisting jet is realised when the flux rope formed during the breakout phase reaches the breakout current sheet. Our jet is a combination of an untwisting flux rope and impulsive interchange reconnection in the flare current sheet formed below the flux rope. Flare loops created by the low-lying reconnection in our model correspond to the jet bright point. The majority of the energy transmitted to the open field of the corona is in the form of a Poynting flux associated with a nonlinear torsional Alfv\'{e}n wave, which is launched by reconnection between the twisted internal flux rope and the untwisted external field. Our findings highlight the similarities between eruptive events across different scales in the solar atmosphere and demonstrates the universality of the breakout mechanism for explaining them \citep{Wyper2017}.

In future work, we aim to assess how our model performs when effects such as gravitational stratification and heating terms are included. Further understanding the magnetic topology and its relation to flare brightening and high-energy particles also are expected to give valuable insight into these events. 

\acknowledgments
This work was supported through PFW's award of a Royal Astronomical Society Fellowship and also through CRD's and SKA's participation in NASA's Living With a Star and Heliophysics Supporting Research programs. Computer resources for the numerical simulations were provided to CRD by NASA's High-End Computing program at the NASA Center for Climate Simulation. We are grateful to J.\ T.\ Karpen, {P.\ Kumar, C.\ E.\ DeForest, N.\ E.\ Raouafi, V.\ M.\ Uritsky, and M.\ A.\ Roberts} for numerous helpful discussions of jets and their observations.


\begin{thebibliography}{}
\expandafter\ifx\csname natexlab\endcsname\relax\def\natexlab#1{#1}\fi

\bibitem[{{Adams} {et~al.}(2014){Adams}, {Sterling}, {Moore}, \&
  {Gary}}]{Adams2014}
{Adams}, M., {Sterling}, A.~C., {Moore}, R.~L., \& {Gary}, G.~A. 2014, \apj,
  783, 11

\bibitem[{{Antiochos}(1998)}]{Antiochos1998}
{Antiochos}, S.~K. 1998, \apjl, 502, L181

\bibitem[{{Antiochos} {et~al.}(1994){Antiochos}, {Dahlburg}, \&
  {Klimchuk}}]{Antiochos1994}
{Antiochos}, S.~K., {Dahlburg}, R.~B., \& {Klimchuk}, J.~A. 1994, \apjl, 420,
  L41

\bibitem[{{Antiochos} {et~al.}(1999){Antiochos}, {DeVore}, \&
  {Klimchuk}}]{Antiochos1999}
{Antiochos}, S.~K., {DeVore}, C.~R., \& {Klimchuk}, J.~A. 1999, \apj, 510, 485

\bibitem[{{Archontis} \& {Hood}(2013)}]{Archontis2013}
{Archontis}, V., \& {Hood}, A.~W. 2013, \apjl, 769, L21

\bibitem[{{Archontis} {et~al.}(2005){Archontis}, {Moreno-Insertis},
  {Galsgaard}, \& {Hood}}]{Archontis2005}
{Archontis}, V., {Moreno-Insertis}, F., {Galsgaard}, K., \& {Hood}, A.~W. 2005,
  \apj, 635, 1299

\bibitem[{{Archontis} \& {T{\"o}r{\"o}k}(2008)}]{Archontis2008}
{Archontis}, V., \& {T{\"o}r{\"o}k}, T. 2008, \aap, 492, L35

\bibitem[{{Archontis} {et~al.}(2010){Archontis}, {Tsinganos}, \&
  {Gontikakis}}]{Archontis2010}
{Archontis}, V., {Tsinganos}, K., \& {Gontikakis}, C. 2010, \aap, 512, L2

\bibitem[{{Aulanier} {et~al.}(2002){Aulanier}, {DeVore}, \&
  {Antiochos}}]{Aulanier2002}
{Aulanier}, G., {DeVore}, C.~R., \& {Antiochos}, S.~K. 2002, \apjl, 567, L97

\bibitem[{{Canfield} {et~al.}(1996){Canfield}, {Reardon}, {Leka}, {Shibata},
  {Yokoyama}, \& {Shimojo}}]{Canfield1996}
{Canfield}, R.~C., {Reardon}, K.~P., {Leka}, K.~D., {et~al.} 1996, \apj, 464,
  1016

\bibitem[{{Chae} {et~al.}(1999){Chae}, {Qiu}, {Wang}, \& {Goode}}]{Chae1999}
{Chae}, J., {Qiu}, J., {Wang}, H., \& {Goode}, P.~R. 1999, \apjl, 513, L75

\bibitem[{{Chandrashekhar} {et~al.}(2014){Chandrashekhar}, {Morton},
  {Banerjee}, \& {Gupta}}]{Chandrashekhar2014}
{Chandrashekhar}, K., {Morton}, R.~J., {Banerjee}, D., \& {Gupta}, G.~R. 2014,
  \aap, 562, A98

\bibitem[{{Cirtain} {et~al.}(2007){Cirtain}, {Golub}, {Lundquist}, {van
  Ballegooijen}, {Savcheva}, {Shimojo}, {DeLuca}, {Tsuneta}, {Sakao}, {Reeves},
  {Weber}, {Kano}, {Narukage}, \& {Shibasaki}}]{Cirtain2007}
{Cirtain}, J.~W., {Golub}, L., {Lundquist}, L., {et~al.} 2007, \sci, 318, 1580

\bibitem[{{Dalmasse} {et~al.}(2012){Dalmasse}, {Pariat}, {Antiochos}, \&
  {DeVore}}]{Dalmasse2012}
{Dalmasse}, K., {Pariat}, E., {Antiochos}, S.~K., \& {DeVore}, C.~R. 2012, in
  EAS Publications Series, Vol.~55, EAS Publications Series, ed.
  M.~{Faurobert}, C.~{Fang}, \& T.~{Corbard}, 201--205

\bibitem[{{DeVore} \& {Antiochos}(2000)}]{DeVore2000}
{DeVore}, C.~R., \& {Antiochos}, S.~K. 2000, \apj, 539, 954

\bibitem[{{DeVore} \& {Antiochos}(2008)}]{DeVore2008}
{DeVore}, C.~R., \& {Antiochos}, S.~K. 2008, \apj, 680, 740

\bibitem[{{Fang} {et~al.}(2014){Fang}, {Fan}, \& {McIntosh}}]{Fang2014}
{Fang}, F., {Fan}, Y., \& {McIntosh}, S.~W. 2014, \apjl, 789, L19

\bibitem[{{Galsgaard} {et~al.}(2005){Galsgaard}, {Moreno-Insertis},
  {Archontis}, \& {Hood}}]{Galsgaard2005}
{Galsgaard}, K., {Moreno-Insertis}, F., {Archontis}, V., \& {Hood}, A. 2005,
  \apjl, 618, L153

\bibitem[{{Gontikakis} {et~al.}(2009){Gontikakis}, {Archontis}, \&
  {Tsinganos}}]{Gontikakis2009}
{Gontikakis}, C., {Archontis}, V., \& {Tsinganos}, K. 2009, \aap, 506, L45

\bibitem[{{Hong} {et~al.}(2014){Hong}, {Jiang}, {Yang}, {Bi}, {Li}, {Yang}, \&
  {Yang}}]{Hong2014}
{Hong}, J., {Jiang}, Y., {Yang}, J., {et~al.} 2014, \apj, 796, 73

\bibitem[{{Hong} {et~al.}(2016){Hong}, {Jiang}, {Yang}, {Yang}, {Xu}, \&
  {Xiang}}]{Hong2016}
{Hong}, J., {Jiang}, Y., {Yang}, J., {et~al.} 2016, \apj, 830, 60

\bibitem[{{Hong} {et~al.}(2011){Hong}, {Jiang}, {Zheng}, {Yang}, {Bi}, \&
  {Yang}}]{Hong2011}
{Hong}, J., {Jiang}, Y., {Zheng}, R., {et~al.} 2011, \apjl, 738, L20

\bibitem[{{Hong} {et~al.}(2013){Hong}, {Jiang}, {Yang}, {Zheng}, {Bi}, {Li},
  {Yang}, \& {Yang}}]{Hong2013}
{Hong}, J.-C., {Jiang}, Y.-C., {Yang}, J.-Y., {et~al.} 2013, \raa, 13, 253

\bibitem[{{Hood} {et~al.}(2012){Hood}, {Archontis}, \& {MacTaggart}}]{Hood2012}
{Hood}, A.~W., {Archontis}, V., \& {MacTaggart}, D. 2012, \solphys, 278, 3

\bibitem[{{Innes} {et~al.}(2009){Innes}, {Genetelli}, {Attie}, \&
  {Potts}}]{Innes2009}
{Innes}, D.~E., {Genetelli}, A., {Attie}, R., \& {Potts}, H.~E. 2009, \aap,
  495, 319

\bibitem[{{Innes} {et~al.}(2010){Innes}, {McIntosh}, \&
  {Pietarila}}]{Innes2010}
{Innes}, D.~E., {McIntosh}, S.~W., \& {Pietarila}, A. 2010, \aap, 517, L7

\bibitem[{{Karpen} {et~al.}(2012){Karpen}, {Antiochos}, \&
  {DeVore}}]{Karpen2012}
{Karpen}, J.~T., {Antiochos}, S.~K., \& {DeVore}, C.~R. 2012, \apj, 760, 81

\bibitem[{{Karpen} {et~al.}(2001){Karpen}, {Antiochos}, {Hohensee}, {Klimchuk},
  \& {MacNeice}}]{Karpen2001}
{Karpen}, J.~T., {Antiochos}, S.~K., {Hohensee}, M., {Klimchuk}, J.~A., \&
  {MacNeice}, P.~J. 2001, \apjl, 553, L85

\bibitem[{{Karpen} {et~al.}(2017){Karpen}, {DeVore}, {Antiochos}, \&
  {Pariat}}]{Karpen2017}
{Karpen}, J.~T., {DeVore}, C.~R., {Antiochos}, S.~K., \& {Pariat}, E. 2017,
  \apj, 834, 62

\bibitem[{{Karpen} {et~al.}(2005){Karpen}, {Tanner}, {Antiochos}, \&
  {DeVore}}]{Karpen2005}
{Karpen}, J.~T., {Tanner}, S.~E.~M., {Antiochos}, S.~K., \& {DeVore}, C.~R.
  2005, \apj, 635, 1319

\bibitem[{{Kliem} \& {T{\"o}r{\"o}k}(2006)}]{Kliem2006}
{Kliem}, B., \& {T{\"o}r{\"o}k}, T. 2006, \prlet, 96, 255002

\bibitem[{{Leake} {et~al.}(2014){Leake}, {Linton}, \& {Antiochos}}]{Leake2014}
{Leake}, J.~E., {Linton}, M.~G., \& {Antiochos}, S.~K. 2014, \apj, 787, 46

\bibitem[{{Liu} {et~al.}(2011){Liu}, {Berger}, {Title}, {Tarbell}, \&
  {Low}}]{Liu2011}
{Liu}, W., {Berger}, T.~E., {Title}, A.~M., {Tarbell}, T.~D., \& {Low}, B.~C.
  2011, \apj, 728, 103

\bibitem[{{Luna} {et~al.}(2012){Luna}, {Karpen}, \& {DeVore}}]{Luna2012}
{Luna}, M., {Karpen}, J.~T., \& {DeVore}, C.~R. 2012, \apj, 746, 30

\bibitem[{{Lynch} {et~al.}(2008){Lynch}, {Antiochos}, {DeVore}, {Luhmann}, \&
  {Zurbuchen}}]{Lynch2008}
{Lynch}, B.~J., {Antiochos}, S.~K., {DeVore}, C.~R., {Luhmann}, J.~G., \&
  {Zurbuchen}, T.~H. 2008, \apj, 683, 1192

\bibitem[{{Lynch} {et~al.}(2009){Lynch}, {Antiochos}, {Li}, {Luhmann}, \&
  {DeVore}}]{Lynch2009}
{Lynch}, B.~J., {Antiochos}, S.~K., {Li}, Y., {Luhmann}, J.~G., \& {DeVore},
  C.~R. 2009, \apj, 697, 1918

\bibitem[{{MacNeice} {et~al.}(2004){MacNeice}, {Antiochos}, {Phillips},
  {Spicer}, {DeVore}, \& {Olson}}]{MacNeice2004}
{MacNeice}, P., {Antiochos}, S.~K., {Phillips}, A., {et~al.} 2004, \apj, 614,
  1028

\bibitem[{{MacNeice} {et~al.}(2000){MacNeice}, {Olson}, {Mobarry}, {de
  Fainchtein}, \& {Packer}}]{MacNeice2000}
{MacNeice}, P., {Olson}, K.~M., {Mobarry}, C., {de Fainchtein}, R., \&
  {Packer}, C. 2000, \cophc, 126, 330

\bibitem[{{Masson} {et~al.}(2013){Masson}, {Antiochos}, \&
  {DeVore}}]{Masson2013}
{Masson}, S., {Antiochos}, S.~K., \& {DeVore}, C.~R. 2013, \apj, 771, 82

\bibitem[{{Masson} {et~al.}(2012){Masson}, {Aulanier}, {Pariat}, \&
  {Klein}}]{Masson2012}
{Masson}, S., {Aulanier}, G., {Pariat}, E., \& {Klein}, K.-L. 2012, \soph, 276,
  199

\bibitem[{{Miyagoshi} \& {Yokoyama}(2003)}]{Miyagoshi2003}
{Miyagoshi}, T., \& {Yokoyama}, T. 2003, \apjl, 593, L133

\bibitem[{{Miyagoshi} \& {Yokoyama}(2004)}]{Miyagoshi2004}
{Miyagoshi}, T., \& {Yokoyama}, T. 2004, \apj, 614, 1042

\bibitem[{{Moore} {et~al.}(2010){Moore}, {Cirtain}, {Sterling}, \&
  {Falconer}}]{Moore2010}
{Moore}, R.~L., {Cirtain}, J.~W., {Sterling}, A.~C., \& {Falconer}, D.~A. 2010,
  \apj, 720, 757

\bibitem[{{Moore} \& {Roumeliotis}(1992)}]{Moore1992}
{Moore}, R.~L., \& {Roumeliotis}, G. 1992, in Lecture Notes in Physics, Berlin
  Springer Verlag, Vol. 399, IAU Colloq. 133: Eruptive Solar Flares, ed.
  Z.~{Svestka}, B.~V. {Jackson}, \& M.~E. {Machado}, 69

\bibitem[{{Moore} {et~al.}(2015){Moore}, {Sterling}, \& {Falconer}}]{Moore2015}
{Moore}, R.~L., {Sterling}, A.~C., \& {Falconer}, D.~A. 2015, \apj, 806, 11

\bibitem[{{Moore} {et~al.}(2001){Moore}, {Sterling}, {Hudson}, \&
  {Lemen}}]{Moore2001}
{Moore}, R.~L., {Sterling}, A.~C., {Hudson}, H.~S., \& {Lemen}, J.~R. 2001,
  \apj, 552, 833

\bibitem[{{Moreno-Insertis} \& {Galsgaard}(2013)}]{Moreno-Insertis2013}
{Moreno-Insertis}, F., \& {Galsgaard}, K. 2013, \apj, 771, 20

\bibitem[{{Moreno-Insertis} {et~al.}(2008){Moreno-Insertis}, {Galsgaard}, \&
  {Ugarte-Urra}}]{Moreno-Insertis2008}
{Moreno-Insertis}, F., {Galsgaard}, K., \& {Ugarte-Urra}, I. 2008, \apjl, 673,
  L211

\bibitem[{{Nistic{\`o}} {et~al.}(2009){Nistic{\`o}}, {Bothmer}, {Patsourakos},
  \& {Zimbardo}}]{Nistico2009}
{Nistic{\`o}}, G., {Bothmer}, V., {Patsourakos}, S., \& {Zimbardo}, G. 2009,
  \solphys, 259, 87

\bibitem[{{Panesar} {et~al.}(2016){Panesar}, {Sterling}, {Moore}, \&
  {Chakrapani}}]{Panesar2016}
{Panesar}, N.~K., {Sterling}, A.~C., {Moore}, R.~L., \& {Chakrapani}, P. 2016,
  \apjl, 832, L7

\bibitem[{{Pariat} {et~al.}(2009){Pariat}, {Antiochos}, \&
  {DeVore}}]{Pariat2009}
{Pariat}, E., {Antiochos}, S.~K., \& {DeVore}, C.~R. 2009, \apj, 691, 61

\bibitem[{{Pariat} {et~al.}(2010){Pariat}, {Antiochos}, \&
  {DeVore}}]{Pariat2010}
{Pariat}, E., {Antiochos}, S.~K., \& {DeVore}, C.~R. 2010, \apj, 714, 1762

\bibitem[{{Pariat} {et~al.}(2015){Pariat}, {Dalmasse}, {DeVore}, {Antiochos},
  \& {Karpen}}]{Pariat2015}
{Pariat}, E., {Dalmasse}, K., {DeVore}, C.~R., {Antiochos}, S.~K., \& {Karpen},
  J.~T. 2015, \aap, 573, A130

\bibitem[{{Pariat} {et~al.}(2016){Pariat}, {Dalmasse}, {DeVore}, {Antiochos},
  \& {Karpen}}]{Pariat2016}
{Pariat}, E., {Dalmasse}, K., {DeVore}, C.~R., {Antiochos}, S.~K., \& {Karpen},
  J.~T. 2016, \aap, 596, A36


\bibitem[{{Patsourakos} {et~al.}(2008){Patsourakos}, {Pariat}, {Vourlidas},
  {Antiochos}, \& {Wuelser}}]{Patsourakos2008}
{Patsourakos}, S., {Pariat}, E., {Vourlidas}, A., {Antiochos}, S.~K., \&
  {Wuelser}, J.~P. 2008, \apjl, 680, L73

\bibitem[{{Phillips} {et~al.}(2005){Phillips}, {MacNeice}, \&
  {Antiochos}}]{Phillips2005}
{Phillips}, A.~D., {MacNeice}, P.~J., \& {Antiochos}, S.~K. 2005, \apjl, 624,
  L129

\bibitem[{{Pontin} {et~al.}(2007){Pontin}, {Bhattacharjee}, \&
  {Galsgaard}}]{Pontin2007}
{Pontin}, D.~I., {Bhattacharjee}, A., \& {Galsgaard}, K. 2007, \phpl, 14,
  052106

\bibitem[{{Raouafi} {et~al.}(2010){Raouafi}, {Georgoulis}, {Rust}, \&
  {Bernasconi}}]{Raouafi2010}
{Raouafi}, N.-E., {Georgoulis}, M.~K., {Rust}, D.~M., \& {Bernasconi}, P.~N.
  2010, \apj, 718, 981
  
\bibitem[{{Raouafi} {et~al.}(2016){Raouafi}, {Patsourakos}, {Pariat}, {Young},
  {Sterling}, {Savcheva}, {Shimojo}, {Moreno-Insertis}, {DeVore}, {Archontis},
  {T{\"o}r{\"o}k}, {Mason}, {Curdt}, {Meyer}, {Dalmasse}, \&
  {Matsui}}]{Raouafi2016}
{Raouafi}, N.~E., {Patsourakos}, S., {Pariat}, E., {et~al.} 2016, \ssr, 201, 1

\bibitem[{{Savcheva} {et~al.}(2007){Savcheva}, {Cirtain}, {Deluca},
  {Lundquist}, {Golub}, {Weber}, {Shimojo}, {Shibasaki}, {Sakao}, {Narukage},
  {Tsuneta}, \& {Kano}}]{Savcheva2007}
{Savcheva}, A., {Cirtain}, J., {Deluca}, E.~E., {et~al.} 2007, \pasj, 59, 771

\bibitem[{{Shen} {et~al.}(2011){Shen}, {Liu}, {Su}, \& {Ibrahim}}]{Shen2011}
{Shen}, Y., {Liu}, Y., {Su}, J., \& {Ibrahim}, A. 2011, \apjl, 735, L43

\bibitem[{{Shibata} \& {Uchida}(1986)}]{Shibata1986}
{Shibata}, K., \& {Uchida}, Y. 1986, \solphys, 103, 299

\bibitem[{{Shibata} {et~al.}(1992){Shibata}, {Ishido}, {Acton}, {Strong},
  {Hirayama}, {Uchida}, {McAllister}, {Matsumoto}, {Tsuneta}, {Shimizu},
  {Hara}, {Sakurai}, {Ichimoto}, {Nishino}, \& {Ogawara}}]{Shibata1992}
{Shibata}, K., {Ishido}, Y., {Acton}, L.~W., {et~al.} 1992, \pasj, 44, L173

\bibitem[{{Shimojo} {et~al.}(1996){Shimojo}, {Hashimoto}, {Shibata},
  {Hirayama}, {Hudson}, \& {Acton}}]{Shimojo1996}
{Shimojo}, M., {Hashimoto}, S., {Shibata}, K., {et~al.} 1996, \pasj, 48, 123

\bibitem[{{Sterling} {et~al.}(2015){Sterling}, {Moore}, {Falconer}, \&
  {Adams}}]{Sterling2015}
{Sterling}, A.~C., {Moore}, R.~L., {Falconer}, D.~A., \& {Adams}, M. 2015,
  Natur, 523, 437

\bibitem[{{Titov}(2007)}]{Titov2007}
{Titov}, V.~S. 2007, \apj, 660, 863

\bibitem[{{Titov} {et~al.}(2002){Titov}, {Hornig}, \&
  {D{\'e}moulin}}]{Titov2002}
{Titov}, V.~S., {Hornig}, G., \& {D{\'e}moulin}, P. 2002, \jgra, 107, 1164

\bibitem[{{T{\"o}r{\"o}k} \& {Kliem}(2005)}]{Torok2005}
{T{\"o}r{\"o}k}, T., \& {Kliem}, B. 2005, \apjl, 630, L97

\bibitem[{{van Ballegooijen} \& {Martens}(1989)}]{vanBallegooijen1989}
{van Ballegooijen}, A.~A., \& {Martens}, P.~C.~H. 1989, \apj, 343, 971

\bibitem[{{Wang} {et~al.}(1998){Wang}, {Sheeley}, {Socker}, {Howard},
  {Brueckner}, {Michels}, {Moses}, {St.~Cyr}, {Llebaria}, \&
  {Delaboudini{\`e}re}}]{Wang1998}
{Wang}, Y.-M., {Sheeley}, Jr., N.~R., {Socker}, D.~G., {et~al.} 1998, \apj,
  508, 899

\bibitem[{{Wyper} {et~al.}(2017){Wyper}, {Antiochos}, \& {DeVore}}]{Wyper2017}
{Wyper}, P.~F., {Antiochos}, S.~K., \& {DeVore}, C.~R. 2017, \nat, 544, 452

\bibitem[{{Wyper} \& {DeVore}(2016)}]{Wyper2016}
{Wyper}, P.~F., \& {DeVore}, C.~R. 2016, \apj, 820, 77

\bibitem[{{Wyper} {et~al.}(2016){Wyper}, {DeVore}, {Karpen}, \&
  {Lynch}}]{Wyper2016b}
{Wyper}, P.~F., {DeVore}, C.~R., {Karpen}, J.~T., \& {Lynch}, B.~J. 2016, \apj,
  827, 4

\bibitem[{{Wyper} \& {Pontin}(2014{\natexlab{a}})}]{Wyper2014a}
{Wyper}, P.~F., \& {Pontin}, D.~I. 2014{\natexlab{a}}, \phpl, 21, 082114

\bibitem[{{Wyper} \& {Pontin}(2014{\natexlab{b}})}]{Wyper2014b}
{Wyper}, P.~F., \& {Pontin}, D.~I. 2014{\natexlab{b}}, \phpl, 21, 102102

\bibitem[{{Yokoyama} \& {Shibata}(1995)}]{Yokoyama1995}
{Yokoyama}, T., \& {Shibata}, K. 1995, Natur, 375, 42

\bibitem[{{Yokoyama} \& {Shibata}(1996)}]{Yokoyama1996}
{Yokoyama}, T., \& {Shibata}, K. 1996, \pasj, 48, 353

\bibitem[{{Young} \& {Muglach}(2014{\natexlab{a}})}]{Young2014a}
{Young}, P.~R., \& {Muglach}, K. 2014{\natexlab{a}}, \pasj, 66, S12

\bibitem[{{Young} \& {Muglach}(2014{\natexlab{b}})}]{Young2014b}
{Young}, P.~R., \& {Muglach}, K. 2014{\natexlab{b}}, \solphys, 289, 3313

\bibitem[{{Zhang} {et~al.}(2016){Zhang}, {Li}, {Ning}, {Su}, {Ji}, \&
  {Guo}}]{Zhang2016b}
{Zhang}, Q.~M., {Li}, D., {Ning}, Z.~J., {et~al.} 2016, \apj, 827, 27

\bibitem[{{Zheng} {et~al.}(2012){Zheng}, {Jiang}, {Yang}, {Bi}, {Hong}, {Yang}, \& {Yang}}]{Zheng2012}
{Zheng}, R., {Jiang}, Y., {Yang}, J., {Bi}, Y., {Hong}, J., {Yang}, D., \& {Yang}, B.\ 2012, \apj, 753, 112

\end{thebibliography}

\end{document}